\documentclass[12pt,a4paper]{article}
\pdfoutput=1

\usepackage{jheppub-arxiv}
\usepackage[dvipsnames,table]{xcolor}
 
\usepackage{amssymb} 
\usepackage{amsmath}
\usepackage{mathtools}
\usepackage{amsfonts}    
\usepackage{dsfont}
\usepackage{pdfpages}
\usepackage{verbatim}
\usepackage{stmaryrd}
\usepackage{graphicx}
\usepackage{import}
\usepackage{etoc}
\usepackage{enumitem}
\usepackage{tensor}
\usepackage{mathrsfs}
\usepackage{textgreek} 
\usepackage[mathscr]{euscript}
\usepackage[smalltableaux]{ytableau}
\ytableausetup{centertableaux}
\usepackage{tikz}
\usetikzlibrary{decorations.pathreplacing, decorations.markings,calc,shapes.misc,decorations.pathmorphing,patterns.meta, math,positioning}
\usepackage{tabularx,ragged2e}
\usepackage{physics}

\usepackage{slashed}
\usepackage{datetime}
\usepackage{hyperref}
\hypersetup{
    pdfencoding=unicode,
	colorlinks=true,
	urlcolor=Maroon,
	linkcolor=RoyalBlue,
	citecolor=Maroon,
	pdftitle={Records from the S-Matrix Marathon: Schwinger-Keldysh Formalism},
	pdfauthor={},
	pdfdisplaydoctitle=true,
	pdfstartview=FitH,
	linktocpage=true
}

\usetikzlibrary{shapes}
\usepackage{enumitem}
\usepackage[export]{adjustbox}
\usepackage{nccmath}
\usepackage{bbm}
\usepackage{braket}
\usepackage{empheq}
\usepackage{cancel}
\usepackage{cleveref}
\usepackage[most]{tcolorbox}

\definecolor{charcoal}{HTML}{343837}

\usepackage{bm}
\definecolor{yellowish}{rgb}{0.880722,0.611041,0.142051}

\newcommand{\ba}{\begin{align}}

\newcommand{\be}{\begin{equation}}
\newcommand{\ee}{\end{equation}}
\def\bd{\begin{tikzpicture}}
\def\ed{\end{tikzpicture}}

\renewcommand\Im{\mathop{\text{Im}}}

\renewcommand\Re{\mathop{\text{Re}}}

\allowdisplaybreaks

\def\XXint#1#2#3{{\setbox0=\hbox{$#1{#2#3}{\int}$}
     \vcenter{\hbox{$#2#3$}}\kern-.5\wd0}}

\definecolor{light-gray}{gray}{0.75}

\renewcommand\d{\text{d}}

\newcommand{\e}{\mathrm{e}}

\newcommand{\G}{\mathcal{G}}

\newcommand{\D}{\mathbb{D}}

\renewcommand{\leq}{\leqslant}
\renewcommand{\geq}{\geqslant}

\definecolor{dpurple}  {RGB} {189,  147,  249}

\usepackage{ntheorem}
\usepackage[framemethod=tikz,footnoteinside=false,hidealllines=true,topline=true,bottomline=true,linecolor=black!10!white,linewidth=1pt,innerleftmargin=0em,innerrightmargin=0em,frametitlerule=true,ntheorem]{mdframed}
\theorembodyfont{\upshape}

\newmdtheoremenv{mtheorem}{Theorem}[section]
\newmdtheoremenv[]{mdexample}{Example}[section]
\newmdtheoremenv{mdremark}{Remark}[section]
\newmdtheoremenv{mddefinition}{Definition}[section]
\newmdtheoremenv{mdcorollary}{Corollary}[section]
\newmdtheoremenv{mdproposition}{Proposition}[section]
\newmdtheoremenv{QA}{Audience question}[section]

\title{{\Large\normalfont Records from the S-Matrix Marathon:}\\ Schwinger--Keldysh Formalism}

\author{{\normalfont Lecturers:}}
\author[1]{Felix~M.~Haehl,}
\author[2]{Mukund~Rangamani}

\affiliation[1]{School of Mathematical Sciences \& STAG Research Centre, University of Southampton, SO17 1BJ, UK}
\affiliation[2]{Center for Quantum Mathematics and Physics (QMAP),\\ Department of Physics \& Astronomy, University of California, Davis, CA 95616, USA}

\abstract{
These notes provide an overview of real-time techniques in quantum field theories and holography. We outline the general rationale and principles underlying the Schwinger-Keldysh formalism and its out-of-time-order generalizations. To illustrate its broad utility we frame our discussion in the language of open quantum dynamics. As a specific application we describe how such real-time observables help in understanding scrambling dynamics. We also describe the holographic prescription for computing thermal Schwinger-Keldysh observables. 

These notes are based on a series of lectures held during the S-Matrix Marathon workshop at the Institute for Advanced Study on 11--22 March 2024.
}

\begin{document}

\maketitle
\setcounter{page}{1}

\setcounter{tocdepth}{4}
\setcounter{secnumdepth}{4}

\makeatletter
\g@addto@macro\bfseries{\boldmath}
\makeatother

\newpage
\section*{Preface}

This article is a chapter from the \emph{Records from the S-Matrix Marathon}, a series of lecture notes covering selected topics on scattering amplitudes~\cite{RecordsBook}. They are based on lectures delivered during a workshop on 11--22 March 2024 at the Institute for Advanced Study in Princeton, NJ. We hope that they can serve as a pedagogical introduction to the topics surrounding the S-matrix theory.

These lecture notes were prepared by the lecturers. The list of references is not meant to be comprehensive, but serve as a guideline for resources we drew from for these lectures. We encourage the interested reader to consult the articles referred herein for a more complete list of references.

\vfill
\section*{Acknowledgments}

It is a pleasure to thank our collaborators Changha Choi, Mark Mezei,  R.~Loganayagam, Moshe Rozali, Gabor Sarosi, Alexandre Streicher, Julio Virrueta, and Ying Zhao, for their insights and collaboration on topics discussed herein. We would especially like to acknowledge our long and fruitful collaboration with R.~Loganayagam.  Much of our understanding of the subject was shaped by extensive discussions on the topic over many years with him.

F.H.~was supported by UK Research and Innovation (UKRI) under the UK government’s
Horizon Europe funding Guarantee EP/X030334/1. M.R.~was supported by U.S.~Department of Energy grant DE-SC0009999 and funds from the University of California. 
The S-Matrix Marathon workshop was sponsored by the Institute for Advanced Study and the Carl P. Feinberg Program in Cross-Disciplinary Innovation.

\setcounter{section}{1}
\newcolumntype{C}{>{\Centering\arraybackslash}X}

\newcommand{\address}[1]{\vbox{\center\em#1}}
\def\D{\mathrm{D}}
\newcommand{\av}{\text{av}}
\newcommand{\dif}{\text{dif}}
\newcommand{\adv}{\text{adv}}
\newcommand{\ret}{\text{ret}}
\renewcommand{\G}{G}
\newcommand{\Gb}{{\bar{G}}}
\NewDocumentCommand{\comt}{>{\SplitArgument{1}{,}}m}{%
  \comtAux#1%
}
\NewDocumentCommand{\comtAux}{mm}{%
  \ensuremath{[#1,#2]}%
}
\NewDocumentCommand{\anti}{>{\SplitArgument{1}{,}}m}{%
  \antiAux#1%
}
\NewDocumentCommand{\antiAux}{mm}{%
  \ensuremath{\{#1,#2\}}%
}
\newcommand{\qmax}{\lfloor\frac{n+1}{2}\rfloor}
\newcommand{\KB}[2]{(#1,#2)_{_\text{SK}}}
\newcommand*{\boxcolor}{black}
\makeatletter
\renewcommand{\boxed}[1]{\textcolor{\boxcolor}{%
\tikz[baseline={([yshift=-1ex]current bounding box.center)}] \node [rectangle, minimum width=1ex,rounded corners,draw] {\normalcolor\m@th$\;\,\displaystyle#1\;\,$};}}
 \makeatother
\usetikzlibrary{decorations.pathmorphing}
\usetikzlibrary{decorations.pathreplacing}
\usetikzlibrary{decorations.markings}
\tikzset{snake it/.style={decorate, decoration=snake}}
\tikzset{->-/.style={decoration={
  markings,
  mark=at position .5 with {\arrow{>}}},postaction={decorate}}}
  
\makeatletter
\newenvironment{sqcases}{
  \matrix@check\sqcases\env@sqcases
}{
  \endarray\right.
}
\def\env@sqcases{
  \let\@ifnextchar\new@ifnextchar
  \left\lbrack
  \def\arraystretch{1.2}
  \array{@{}l@{\quad}l@{}}
}
\makeatother

\newpage

\section*{\label{ch:RangamaniHaehl}Schwinger--Keldysh Formalism\\
\normalfont{\textit{Felix Haehl, Mukund Rangamani}}}

\setcounter{section}{0}

\noindent\rule{\textwidth}{0.25pt}
\vspace{-0.8em}
\etocsettocstyle{\noindent\textbf{Contents}\vskip0pt}{}
\localtableofcontents
\vspace{0.5em}
\noindent\rule{\textwidth}{0.25pt}
\vspace{1em}

\section[Invitation to the Schwinger--Keldysh path integral]{Invitation to the Schwinger--Keldysh path integral\\
\normalfont{\textit{Mukund Rangamani}}}
\label{secReview}

The Schwinger-Keldysh formalism is broadly aimed at facilitating the computation of real-time observables in quantum systems. In the following, we give a broad physical motivation for it, before diving into the details. The discussion below largely follows the presentation in \cite{Chou:1984es} and is also reviewed in~\cite{Haehl:2016pec}. 

 Let us consider computing in a QFT the two-point Green's function for some (generically complex) Heisenberg operator $\mathcal{O}(x)$ in some pure state
\begin{align}
 G(x,x') = -i \matrixelement{\Omega}{\mathcal{T}\, \mathcal{O}(x)\, \mathcal{O}(x')}{\Omega}\,,
\end{align}
where $\mathcal{T}$ is the standard time ordering, and $x = (t, \mathbf{x})$ is the spacetime coordinate. We will distinguish the temporal direction (coordinatized by $t$) when necessary. The state 
$\ket{\Omega}$ is the ground state of the full interacting theory. In perturbation theory, working in the interaction picture, the non-trivial part of the evolution operator
\begin{equation}
U(t_0,t) = \mathcal{T} \, \exp \left[ -i \int_{t_0}^t \d t' \, H_{\text{int}}(t')\right]\,,
\end{equation}
defines temporal evolution of the interaction picture states. Using this expansion, one then finds an expression for the two-point Green's function:
\begin{equation}\label{eq:Gcalc}
\begin{split}
 G(x,x') 
 = -i \matrixelement{0}{S^\dagger\,  \mathcal{T}\,  {\cal O}(x)\, {\cal O}^\dagger(x') \, S}{0} 
 = -i \frac{\matrixelement{0}{\mathcal{T}[S \; {\cal O}(x) \, {\cal O}^\dagger(x')]}{0} }{\matrixelement{0}{S}{0} }\,.
\end{split}
\end{equation}
We arrive at this equation by introducing the $S$-matrix  $S \equiv U(-\infty, \infty)$ and obtaining the interacting ground state by evolving the non-interacting ground state $\ket{0}$  of the free Hamiltonian $H_0$. We usually use the final expression in~\eqref{eq:Gcalc} as the starting point for perturbation theory.

In writing the second equality, we have expressed the instantaneous late time ground state in terms of the early time state, assuming an adiabatic evolution of the system expressed as a property of the $S$-matrix. Namely, the phase picked up by acting on the final state with $S^\dagger $ is the same as the one accumulated during the evolution, i.e.,
$\bra{0} S^\dagger = \bra{0} \e^{-i\alpha}$ and $ \matrixelement{0}{S}{0} =\e^{i\alpha}$, for some phase $\alpha$. Thus, we assume that the physical content of the ground state remains unchanged during the evolution, up to a phase rotation. This fails in non-equilibrium situations, where adiabatic evolution is not justified.

\begin{figure}[t!]
\begin{center}
\includegraphics[width=.7\textwidth]{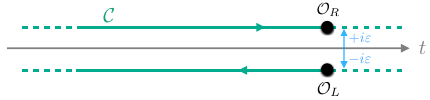}
\caption{An illustration of the generic Schwinger--Keldysh complex time contour. Every operator $\mathcal{O}$ in the original theory has two representations in the Schwinger--Keldysh path integral, viz., $\mathcal{O}_R$ and $\mathcal{O}_L$. The labels help distinguish which  part of the contour the operator is inserted on. There is a natural contour ordering prescription wherein right operators are time-ordered and left operators are anti-time ordered.}
\label{fig:contour1}
\end{center}
\end{figure}

The Schwinger--Keldysh (SK) formalism deals with non-equilibrium dynamics by only ever making reference to the initial state,\footnote{ Hence, the Schwinger--Keldysh formalism is sometimes also referred to as {\it in-in formalism}.} which may be taken without loss of generality to be an equilibrium configuration, the instantaneous vacuum state of $H_0$ at $t=-\infty$. As we evolve the system, we want to ensure that we make no assumption about where (or what state) it would end up at late times. To this end, rather than continuing to evolve forward, we should revert, after allowing the interactions to influence the system, to the initial state. In a path integral, this can be done systematically by introducing a SK-evolution operator, which evolves the system in a complex time contour. Let $\mathcal{C}$ be a contour in the complex time plane that starts out at $t= -\infty+i\varepsilon$, follows the real axis, and then retraces its trajectory back with a small imaginary displacement by $-2i\varepsilon$, cf. Fig.~\ref{fig:contour1}. We have chosen to orient the contour so that the direction of traversal is clockwise (about the origin, say). We will also find it useful to label the forward leg of the contour as the right ($\text{R}$) part and the backward leg the left ($\text{L}$) part.

Given such a contour, we can work with operators which live in this complexified domain, and define the Schwinger--Keldysh S-matrix by working with contour-ordering prescription, to wit,
\begin{equation}
U_\mathcal{C}\equiv \mathcal{T}_\mathcal{C} \, \exp\left[ -i \int_\mathcal{C} \,\d t' \, H_{\text{int}}(t') \right] \,.
\end{equation}
There is a sensible time-ordering prescription inherited from this contour-ordering.

It is often, however, useful not to work with a single contour, but rather, work with fields and operators labeled by which part of the contour they appear on. This perspective is best served by doubling of the degrees of freedom and using it to obtain physical results. We have left and right fields indexed by their position on the Schwinger--Keldysh contour $\mathcal{C}$. Furthermore, as illustrated, the operators on the right/forward leg are time-ordered, those on the left/backward leg are anti-time-ordered, and the right operators precede those on the left leg of the contour.

\begin{figure}[t!]
\begin{center}
\includegraphics[width=.65\textwidth]{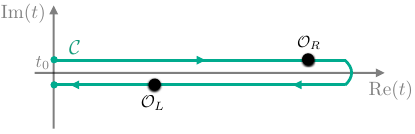}
\caption{SK time contour in the case where the initial state at time $t_0$ is known and the latest operator insertion happens at time $t$. The indicated operator insertions correspond to a real-time correlator $G_<(x,x')$.}
\label{fig:contour2}
\end{center}
\end{figure}

If we have complete knowledge of the state (which we can take  in general to be a density matrix) of the system at some finite time $t_0$, then we do not need to follow the contour all the way from $t=-\infty$ to $+\infty$ and back. It suffices to focus solely on the part of the contour from $t_0$ to $\text{max}(t,t')$, which corresponds to the future-most operator insertion, before retracing back to the initial configuration, cf.~Fig.~\ref{fig:contour2}. Intuitively, all this is saying is that the knowledge of the density matrix can be treated as initial conditions for the subsequent evolution. Furthermore, for finite time computations, details of how the system evolves to the future of all operator insertions (or measurements) are inessential.

%

\subsection{The Schwinger--Keldysh path integral}\label{sec:genis}

Based on the discussion above, we can motivate the general Schwinger--Keldysh path integral,
\begin{equation}\label{eq:skgen}
    {\cal Z}_\text{SK}[J_R , J_L] = \text{Tr} \left[U[J_R] \,\rho_0\,U^\dagger[J_L]  \right] ,
\end{equation}
where R and L refer to the evolution of the ket and bra part of the density matrix.
This computes SK-path ordered correlators, which consist of a product of time-ordered operator insertions followed by a product of anti-time-ordered ones. We denote an operator inserted on contour segment $I \in \{R,L\}$ as $\mathcal{O}_I$. A typical SK correlator is of the form\footnote{Of course, the operator insertions do not all have to be the same. We have adopted a succinct notation, suppressing additional labels which would serve to distinguish operators.}
\begin{equation}
\begin{split}
    \big\langle {\cal T}_{\cal C} \, \mathcal{O}_R(t_1)& \cdots \mathcal{O}_R(t_k) \mathcal{O}_L(t_{k+1}) \cdots \mathcal{O}_L(t_n) \big\rangle
\\&=
    \text{Tr}\left[ \big(\bar{\cal T}\, \mathcal{O}(t_n) \cdots \mathcal{O}(t_{k+1}) \big) \big( {\cal T} \, \mathcal{O}(t_k) \cdots \mathcal{O}(t_1) \big)\,\rho_0\right]\,,
\end{split}
\end{equation}
where ${\cal T}_{\cal C}$, ${\cal T}$, $\bar{\cal T}$ denote contour ordering, time ordering, and anti-time ordering, respectively. The R operators, which correspond to the ket part of $\rho$, are time-ordered, while the L operators, which correspond to the bra part, are anti-time-ordered.

As presaged, the original motivation for the Schwinger--Keldysh formalism is to describe systems interacting with an environment. In this context, it is natural to work with variables that evolve classically, and those that are subject to quantum fluctuations from the environment. This is achieved by switching to the average/difference basis, where difference fields (quantum) parametrize fluctuations:
\begin{equation}
    \mathcal{O}_\av = \frac{\mathcal{O}_R + \mathcal{O}_L}{2} \,,\qquad \mathcal{O}_\dif = \mathcal{O}_R - \mathcal{O}_L \,.
\end{equation}

In this basis, the so-called {\it largest time equation} is
\begin{equation}
\label{eq:largesttime}
    \langle \mathcal{T}_{\cal C} \, {\color{RoyalBlue}\mathcal{O}_\dif(t_f)}\, \mathcal{O}_{I_1}(t_1) \cdots \mathcal{O}_{I_n}(t_n) \rangle = 0 \qquad (t_f > t_1,\ldots , t_n) \,,
\end{equation}
where $\mathcal{O}_{I_i}$ are arbitrary operator combinations on the contours ($I_i=R,L$ or $I_i=\{\text{av},\text{dif}\}$ label the contour segments), and ${\cal T}_{\cal C}$ denotes Schwinger--Keldysh contour ordering. These constraints follow from unitarity: $U^\dagger[J]U[J] = 1$ when $J_R=J_L\equiv J$ since $\mathcal{O}_\av$ is sourced by this $J$.
Pictorially:
\begin{equation*}
\includegraphics[width=\textwidth]{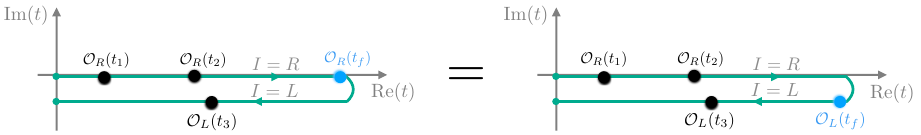}
\end{equation*}
Generally, the coupling between sources and operators is $J_R \, \mathcal{O}_R - J_L \, \mathcal{O}_L = J_\av \,\mathcal{O}_\dif + J_\dif\, \mathcal{O}_\av$.

\subparagraph{Largest time sum rule:} The natural objects on the Schwinger--Keldysh contour are the contour-ordered correlators. Consider first the Schwinger--Keldysh two-point function matrix $\mathbb{G}_{IJ}(x,x') = \langle {\cal T}_{\cal C} \, \mathcal{O}_{I}(x) \mathcal{O}^\dagger_{J}(x') \rangle$, where $I,J\in \{R,L\}$. These can be assembled into a matrix
\begin{equation}
    \mathbb{G}(x,x') = \begin{pmatrix}
        iG_F(x,x') & iG_>(x,x') \\
       i G_<(x,x') & iG_{\bar F}(x,x')
    \end{pmatrix}
    =
    \begin{pmatrix}
        \langle {\cal T} \, \mathcal{O}(x) \mathcal{O}^\dagger(x') \rangle
        &
        \langle   \mathcal{O}^\dagger(x') \mathcal{O}(x)\rangle 
        \\
        \langle \mathcal{O}(x)  \mathcal{O}^\dagger(x') \rangle
        & 
         \langle \overline{\cal T} \, \mathcal{O}(x) \mathcal{O}^\dagger(x') \rangle
    \end{pmatrix}\,.
\end{equation}
In our earlier language, 
\begin{equation}
\begin{split}
G_F(x,x') 
&= 
    -i \matrixelement{\Omega}{\mathcal{T}\, \mathcal{O}(x)\, \mathcal{O}(x')}{\Omega}\,, \\ 
G_{\overline{F}}(x,x') 
&= 
    -i \matrixelement{\Omega}{ \overline{\mathcal{T}}\, \mathcal{O}(x)\, \mathcal{O}(x')}{\Omega}\,, \\ 
G_<(x,x') 
&= 
    -i \matrixelement{\Omega}{\mathcal{O}(x')\, \mathcal{O}(x)}{\Omega}\,, \\ 
G_>(x,x') 
&= 
    -i \matrixelement{\Omega}{\mathcal{O}(x)\, \mathcal{O}(x')}{\Omega}\,. \\ 
\end{split}
\end{equation}
These satisfy the sum rule
\begin{equation}
    G_F + G_{\bar F} - G_> - G_< = 0 \,.
\end{equation}
This sum rule encapsulates once again the unitarity and causality constraints. It is the simplest example of a more general statement. Consider all contour-ordered $n$-point functions; one can readily check the following identity:\footnote{ Here, $I_i = 1,2$ correspond to labels $R,L$.}
\begin{equation}
\label{eq:sumrule}
    \sum_{I_1,\ldots,I_n \in\{1,2\}} (-1)^{I_1 + \ldots + I_n} \;\langle {\cal T}_{\cal C} \, \mathcal{O}_{I_1}(x_1) \cdots \mathcal{O}_{I_n}(x_n) \rangle = 0 \,.
\end{equation}
The cancellation in this sum occurs pairwise: assume without loss of generality that $t_1 > t_2,\ldots, t_n$. Then for every choice of $I_2,\ldots,I_n$ there are two different choices for $I_1$. The difference between these two choices is zero thanks to the largest time equation \eqref{eq:largesttime}.
We will refer to \eqref{eq:sumrule} as the {\it largest time sum rule}.

Given the SK generating function~\eqref{eq:skgen}, one can compute all contour-ordered correlators by varying with respect to the sources $J_R$ and $J_L$. Note that this is a highly redundant picture since there are a priori $2^n$ contour ordered $n$-point functions, which is greater than the number of Wightman functions (which number $n!$). Some of these redundancies are accounted for by the largest time equations. In fact, a useful way to encode this is to invoke a BRST symmetry of the Schwinger-Keldysh construction, see the Appendix for a brief sketch of this idea. Before we turn to finding a simple algorithm to map the contour-ordered observables to physical ones, it is useful to consider a generalization. 

\subsection{The OTO contours}\label{sec:oto}

\begin{figure}[t!]
\centering
\includegraphics[width=.6\textwidth]{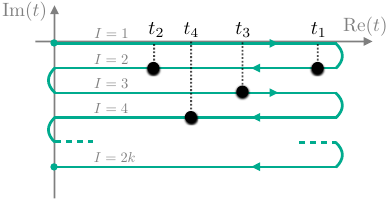}
 \caption{The k-OTO contour, computing the out-of-time-ordered correlation functions encoded in the generating functional \eqref{eq:koto}. As in Fig.~\ref{fig:contour1} and Fig.~\ref{fig:contour2} the physical time runs from left to right, and the vertical excursions are infinitesimal ($i\varepsilon$). We also show insertion points required to compute the correlator with temporal ordering $t_1>t_2$, $t_2<t_3$, and $t_3>t_4$.
}
\label{fig:koto}
\end{figure}

Consider a generic $n$-point function of Heisenberg operators, $\mathcal{O}^{(i)}(t_i)$, that is, the Wightman correlation functions $\expval{\mathcal{O}^{(1)} (t_1) \, \mathcal{O}^{(2)} (t_2)\, \cdots \, \mathcal{O}^{(n)} (t_n) }$, without prescribed temporal ordering. Writing  out this correlation function in terms of Schr\"odinger operators,
 $\mathcal{O}^{(i)}(t_0)$,  using $\mathcal{O}(t) = U(t_0,t)^\dagger \,\mathcal{O}(t_0) \,U(t_0,t)$, we obtain
\begin{equation}\label{eq:Wightman}
\begin{split}
&G(t_1,\cdots, t_n) 
 \equiv
  \expval{\mathcal{O}^{(1)} (t_1) \, \mathcal{O}^{(2)} (t_2)\, \cdots \, \mathcal{O}^{(n)} (t_n) } \\
&\quad = 
    \expval{U^\dagger(t_0,t_1) \, \mathcal{O}^{(1)}(t_0)\, U(t_0,t_1)  \;
    \cdots U^\dagger(t_0,t_n) \,\mathcal{O}^{(n)}(t_0) \,U(t_0,t_n)} .
\end{split}
\end{equation}
The temporal evolution of the system between the operator insertions involves a series of forward and backward evolutions by $U(t_0,t_i)$ and $U^\dagger(t_0,t_j)$, respectively. This is an inevitable consequence of the lack of any temporal ordering.  One can represent such an evolution by a path integral contour, which involves a series of temporal switchbacks; see Fig.~\ref{fig:koto}.

We will refer to such contours as the \emph{timefold} or \emph{out-of-time-order} (OTO) contours. It is useful to decorate the latter with a depth label, $k$, with a $k$-OTO contour specifying a contour, with $k$ forward and $k$ backward evolutions. The Schwinger--Keldysh contour is then a special case, being the $1$-OTO contour. 

The generating function for the $k$-OTO contour can be computed with sources $J$ for the operators $\mathcal{O}$, and encoded in the evolution operator $U[J]$
\begin{equation}
U[J] =  \mathcal{T} \, \exp\left[  -i \int_{t_i}^{t}\, dt\, H[J]   \right] \,, \qquad
(U[J] )^\dagger=  \bar{{\cal T}} \, \exp\left[  i \int_{t_i}^{t}\, dt\, H[J]   \right] \,.
\end{equation}
The $k$-OTO generating function is then defined to be 
\begin{equation}\label{eq:koto}
\mathcal{Z}_{\text{$k$-oto}}[J_{I}]=
\Tr[   \cdots
U[J_{3}]
(U[J_{2}])^\dag  U[J_{1}] \;\rho_i \; (U[J_{2k}])^\dag
U[J_{2k-1}] (U[J_{2k-2}])^\dag \cdots] \,.
\end{equation}
The contour prescription with the decorated labels $I=1,\ldots, 2k$ is useful, but highly redundant. There are several useful bases that are adapted to obtain physical results more directly~\cite{Haehl:2017qfl}.  
We describe them below, with fixed time order, $t_1 > t_2 > \cdots > t_n$, for definiteness:

\begin{itemize}[label=$\diamond$]
\item Wightman basis: This comprises the  $n!$  physical correlators 
\begin{equation}
G_\sigma(t_1, t_2,\, \cdots\,, t_n) =
\expval{\mathcal{O}^{(\sigma_1)} \, \mathcal{O}^{(\sigma_2)}   \, \cdots \, \mathcal{O}^{(\sigma_n)} } \,, \qquad \sigma \in S_n \,,
\label{eq:tobasis}
\end{equation}
where $S_n$ denotes the group of permutations of $n$ objects. 

\item Nested correlators: Another useful set uses nested commutators and anti-commutators of the $n$ operators $\mathcal{O}^{(i)}$. These are constructed in terms of the elementary building blocks, which are commutators $\comt{\cdot,\cdot}$ and anti-commutators $\anti{\cdot,\cdot}$ of the operators. We can consider nesting a sequence of graded commutator, anti-commutators; for example,
\begin{equation}
\comt{\anti{\comt{\mathcal{O}^{(1)},\mathcal{O}^{(2)}},\mathcal{O}^{(3)}},\cdots}\,,
\label{eq:nestelem}
\end{equation}
which illustrates the general idea. This set is spanned by $2^{n-2}n!$ correlators. These objects, together with appropriate time-ordering step functions, form the basis of time-ordered response functions \cite{Chou:1984es} (also see \cite{Haehl:2016pec}). This statement should be familiar for 2-point functions, since the complete information of the propagator is contained in the commutator and anti-commutator. The reason for their importance can be traced to the fact that Lorentzian causal ordering ensures that the commutator of operators will vanish when the insertions are spacelike separated.

\item The LR correlators: The space of correlation functions derived from the $k$-OTO contour forms another basis. There are a total of $(2k)^n$ correlation functions, since $\mathcal{O}(t_i)$ can be inserted in any one of the $2k$ legs of the contour. This vast over-determination can be collapsed to something simpler, for instance, by switching off or aligning some sources, we can collapse some timefolds using unitarity.

\item  The Av-Dif correlators: A closely related basis involves rotating the LR-basis into the average-difference operator basis. This is done by a natural extension of the Keldysh basis used in the usual Schwinger--Keldysh formalism $\mathcal{O}_{\av(\ell)} = \frac{1}{2}\, (\mathcal{O}_{2\ell-1} + \mathcal{O}_{2\ell})$ and $\mathcal{O}_{\dif(\ell)}  = \mathcal{O}_{2\ell-1} - \mathcal{O}_{2\ell}$, where $\ell = 1 ,\ldots,k$.
\end{itemize}

Note that in situations where the state is analytic, we may use an $i\varepsilon$ prescription to recover the Wightman function from the Euclidean Schwinger functions, via 
\begin{equation}
G_\sigma(t_1,t_2,\cdots,t_n) = \lim_{\varepsilon_i \to 0} G_E(\tau_1,\tau_2, \cdots , \tau_n) \big|_{\tau_i = i\, t_i + \varepsilon_i} \qquad (\varepsilon_{\sigma_1} > \ldots > \varepsilon_{\sigma_n})\,.
\end{equation}
This can be implemented directly in 2d CFTs thanks to the convergence of the OPE. A nice discussion of higher-dimensional CFTs can be found in~\cite{Qiao:2022xzc}.

Let us record some general results about these OTO contours:
\begin{itemize}[label=$\diamond$]
\item  The computation of $n$-point Wightman functions can be done on a $k$-OTO contour with $k \leq \qmax$. This can be seen by noting that the worst-case scenario is an ordering in which we first act with $\mathcal{O}(t_1)$, then by $\mathcal{O}(t_n)$, followed by $\mathcal{O}(t_2)$,and then acting with $\mathcal{O}(t_{n-1})$ and so on, involving switching back between forward and backward evolutions.
\item One can trivially embed an $n$-point function on a $k > \qmax $-OTO contour. This will involve some redundancies, but it is useful to derive relations between correlators.
\item For completeness, let us record that all two-point functions can be computed on a SK/1-OTO contour. Of the $6$ three-point functions, $4$ are computed on the 1-OTO contour, but two others $\expval{\mathcal{O}(t_1)\,\mathcal{O}(t_3)\,\mathcal{O}(t_2)}$, and $\expval{\mathcal{O}(t_2)\,\mathcal{O}(t_3)\,\mathcal{O}(t_1)}$ require the use of a $2$-OTO contour. Similarly, only $8$ of the $24$ four-point functions can be computed on the $1$-OTO contour; the remaining $16$ require a 2-OTO contour. 
\end{itemize}

The redundancies on the $k$-OTO contour can be understood pictorially by thinking of the contour as a wireframe of an abacus, with the operators as beads. The operators can slide up and down to occupy their temporal spot on any of the legs, as long as other operators do not obstruct them. The basic reason for this is again the analog of the largest time equation, which originates from the normalization of the time-ordering step function. We define  
\begin{equation}
\Theta_{abcd\cdots} = \Theta_{ab}\, \Theta_{bc}\, \Theta_{cd}\cdots \,, \qquad 
\Theta_{ab} = 1\,, \; \text{ if } t_a > t_b \,.
\end{equation}
From here it follows that, 
\begin{equation}
\sum_{\sigma \in S_n} \Theta_{\sigma_1 \sigma_2 \cdots \sigma_n} =1\,.
\end{equation}
As noted above, this implies that difference operators cannot be future most. This observation can be used to give a simple mapping between the different bases described above using the \emph{Keldysh bracket}. We will illustrate it below for the 1-OTO case, and refer to~\cite{Haehl:2017qfl} for the $k$-OTO generalization.

The Keldysh bracket is defined as an operation that takes as input a physical single copy (non-Schwinger--Keldysh) operator as its first entry, a Schwinger-Keldysh operator as the second entry, and outputs a product of the single copy operators as the result. Letting $I$ be the identity operator, one defines 
\begin{equation}\label{eq:kelbrk}
\begin{aligned}
\KB{A}{B_R} &\equiv A\,B \,,  &\qquad   \KB{A}{B_L} &\equiv  B\,A \,, \\ 
\KB{A}{B_\av} &\equiv \anti{A,B} \,,  
&\qquad  \KB{A}{B_\dif} &\equiv  \comt{A,B} \,,\\
\KB{I}{B_\av} &\equiv B \,,  &\qquad   \KB{I}{B_\dif} &\equiv  0 \,,
\end{aligned}
\end{equation}

A Schwinger--Keldysh correlator can be expressed as a nested Keldysh bracket acting on the identity operator and then applying Schwinger--Keldysh time-ordering. The time-ordering is a particular choice of the step functions. One simply sums over all possible orderings of operators inside the nested Keldysh brackets and dresses each of them with the appropriate causal step function. Specifically, 
\begin{equation}
\begin{split}
& \langle \mathcal{T}_{SK} \ \mathcal{O}^{(1)}_{I_1} \mathcal{O}^{(2)}_{I_2} \ldots \mathcal{O}^{(p)}_{I_p} \rangle \\
& \qquad = \sum_{\substack{\text{time}\\\text{orderings}}} \Theta_{\sigma_1 \sigma_2  \cdots  \sigma_p}\;
  \langle  \KB{\cdots\KB{\KB{I }{ \mathcal{O}^{(\sigma_1)} } }{\mathcal{O}^{(\sigma_2)}},  \cdots}{\mathcal{O}^{(\sigma_p)}}
  \rangle  \,,
 \end{split}
 \label{eq:KBForm}
\end{equation}
where $\sigma_1 \sigma_2\,\cdots \sigma_p$ is a permutation of the $p$ indices. This expression can be used to bring any correlation function to a standard form involving commutators and anti-commutators. For instance, one can check that 
\begin{equation}
\begin{split}
\expval{\mathcal{T}_{SK} A_\av\, B_\av}
&=
	\Theta_{AB}  \, \expval{\KB{ \KB{I}{A_\av}}{B_\av}}
	+ \Theta_{BA} \, \expval{\KB{\KB{I}{B_\av}}{A_\av}} \\
&=
    \Theta_{AB}\, \expval{\KB{A}{B_\av}} + \Theta_{BA}\, \expval{\KB{B}{A_\av}}
= \expval{\anti{A,B}} \,.
\end{split}
\label{eq:2ptaa}
\end{equation}
Similarly, we may further deduce that
\begin{equation}
\begin{split}
\expval{\mathcal{T}_{SK} A_\av\, B_\dif} 
&= 
    \Theta_{AB}\, \expval{\comt{A,B}} \\ 
\expval{\mathcal{T}_{SK} A_\av\, B_\dif\, C_\dif}
&= 
    \Theta_{ABC}\, \expval{\comt{\comt{A,B},C}}  + \Theta_{CAB}\, \expval{\comt{\comt{C,A},B}}\,.
\end{split}
\end{equation}
\\

Let us take stock: the space of $n$-point correlation functions is, in general, spanned by $n!$ distinct Wightman functions. These can be represented on OTO contours in many equivalent ways. To classify them, we can denote as $g_{n,q}$ the number of {\it proper} $q$-OTO $n$-point functions, i.e., those which require at least a $q$-OTO path integral to be representable. A proper $q$-OTO $n$-point function can be represented on a $k$-OTO contour (with $k\geq q$) in some number of ways, $h_{n,k}^{(q)}$. This leads to a decomposition of the number of $n$-point functions into equivalence classes under unitarity rules:
\begin{equation}
n! = \sum_{q=1}^{\lfloor \frac{n+1}{2} \rfloor} g_{n,q} \,,\qquad (2k)^n = \sum_{q=1}^{\lfloor\frac{n+1}{2}\rfloor} g_{n,q} \, h_{n,k}^{(q)} \,,
\label{eq:OTOclassification}
\end{equation}
where we recall that $(2k)^n$ is the (vastly redundant) total number of $n$-point functions representable on the $k$-OTO contour: each of the $n$ operators can be inserted on any of the $2k$ contour segments. The unitarity rules translate into a combinatorial counting problem, which allows us to compute the integers $g_{n,q}$ and $h_{n,k}^{(q)}$ recursively. We refer to \cite{Haehl:2017qfl} for details and results.

\section[Schwinger--Keldysh and Cutkosky cutting rules]{Schwinger--Keldysh and Cutkosky cutting rules\\
\normalfont{\textit{Mukund Rangamani}}}\label{sec:cutkosky}

We will now briefly recall the cutting rules and perturbative unitarity, showing that this formalism is isomorphic to the Schwinger--Keldysh approach described above, cf.~\cite{tHooft:1973wag} for the original discussion. 

For concreteness, consider Feynman diagrams built out of propagators
\begin{equation}
    \begin{split}
    G_F(x) &= \Theta(x^0) G_>(x) + \Theta(-x^0) G_<(x)\,, \\
    G_{\gtrless}(x) &= \int \frac{\d^4k}{(2\pi)^4} \, \delta(k^2-m^2)\, \Theta(\pm k^0)\, \e^{-ikx}\,.
    \end{split}
\end{equation}
Here $G_{\gtrless}(x)$ are the two Wightman functions and refer to the positive and negative frequency parts of the Feynman propagator. Decomposing Feynman diagrams into positive and negative frequency components leads to an efficient way of tracking causality constraints. One introduces a notation where every vertex in a diagram is either circled (black) or uncircled (white) with the following rules for propagators connecting the various vertex combinations:
\begin{equation}
\begin{split}
\includegraphics[width=.07\textwidth]{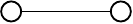} \; &= G_F(x-x')\,, \\
\includegraphics[width=.07\textwidth]{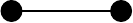} \; &= G_{\bar F}(x-x')\,,\\
\includegraphics[width=.07\textwidth]{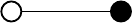} \;  &= G_>(x-x')\,, \\
\includegraphics[width=.07\textwidth]{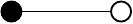} \; &= G_<(x-x')\,.
\end{split}
\end{equation}
In addition, every circled vertex receives a minus sign.
Physically, positive energy is always transferred from uncircled to circled vertices. Clearly, the effective doubling of degrees of freedom (by means of doubling every vertex) is very much analogous to the Schwinger--Keldysh approach. Let us discuss this in more detail.

The largest time sum rule is now the statement that the sum over all circlings of a given diagram vanishes. For example: 
\begin{equation}
\begin{aligned}
\includegraphics[width=.55\textwidth]{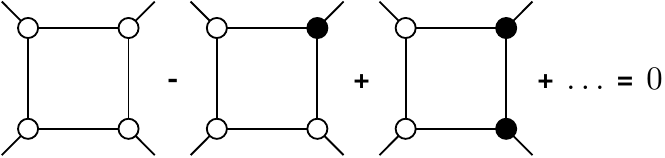}
\end{aligned}
\end{equation}
The cancellation again happens in pairs: any given diagram cancels against its `largest time partner' diagram, i.e., the diagram obtained by changing the circle on the future-most vertex.

Note that generally many diagrams vanish because of the condition that energy can only flow into circled vertices. For instance, an isolated circled vertex that is not connected to any external legs must give a vanishing contribution as energy cannot flow to it from anywhere:
\begin{equation}
\begin{aligned}
\includegraphics[width=.2\textwidth]{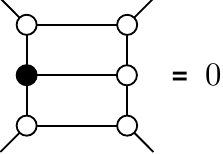}
\end{aligned}
\end{equation}
This constraint is obvious in momentum space, where the diagram is proportional to a vanishing product of $\Theta(\pm k_{ij}^0)$. This makes it less transparent in Schwinger--Keldysh formalism.

We can obtain a more transparent statement by considering again the sum over all circlings in frequency space. Every choice of circlings can be associated with a cut that separates the diagram into a shaded (circled) and sunny (uncircled) part. Energy flows from the unshaded region into the shaded region, and both regions must be connected to external legs in a way that is consistent with this (otherwise the diagram vanishes). For instance:
\begin{equation}
\begin{aligned}
\includegraphics[width=.3\textwidth]{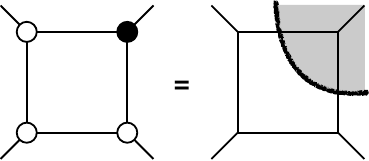}
\end{aligned}
\end{equation}
with the understanding that we put all cut internal propagators on-shell,
\begin{equation}
  \frac{1}{k^2-m^2}  \; \longrightarrow \;  \delta(k^2-m^2) \,,    
\end{equation}
thus turning the cut diagram into a product of amplitudes.
The {\it Cutkosky cutting rule} is precisely the largest time sum rule written as a sum over cuts:
\begin{equation}
    F(k_1,\ldots,k_n) + \bar{F}(k_1,\ldots,k_n) = - \sum_\text{cuts} F_\text{cut}(k_1,\ldots,k_n) \,,
    \label{eq:cutkosky}
\end{equation}
where $F$ is a (momentum space) diagram with all vertices uncircled, $\bar{F}$ is the corresponding diagram with all vertices circled, and $F_\text{cut}$ are the various cut versions of the same diagram.

We conclude this discussion by making the connection with the S-matrix manifest. Recall that unitarity of $S=1+iT$ implies 
\begin{equation}
\label{eq:Tunitarity}
    T - T^\dagger = iT^\dagger T \,.
\end{equation}
The cutting rule \eqref{eq:cutkosky} looks precisely like the contribution of a given diagram $F$ to this unitarity condition, where $T$ accounts for sunny amplitudes and $T^\dagger$ for shaded ones. This is a little too fast, but it does lead us in the right direction: in particular, in order for \eqref{eq:cutkosky} to really encode \eqref{eq:Tunitarity}, we need that the Lagrangian (which defines the Feynman rules for $F$) satisfies ${\cal L} = {\cal L}^\dagger$. This will then ensure that the Feynman rules used to build a diagram $\bar F$ do indeed correspond to those associated with $S^\dagger$ (or $T^\dagger$).

\section[The open quantum system paradigm]{The open quantum system paradigm\\
\normalfont{\textit{Mukund Rangamani}}}\label{sec:oqs}

As described above, the Schwinger--Keldysh formalism and its $k$-OTO generalization provide a framework for computing real-time observables in QFT. One concrete application of these ideas is in the context of non-equilibrium dynamics. We will frame this application in terms of the open quantum system paradigm, wherein some quantum degrees of freedom are traced over. 

Consider a QFT with some degrees of freedom encoded in local fields $\Psi(x,t)$,  which is the physical system of interest. We will imagine this system coupled to an environment, which itself is another field theory, perhaps with many degrees of freedom $X_i$. The unitary microscopic theory is of the form:
\begin{equation}
S[\Psi, X_i] = S_s[\Psi] +  S_e[X_i] + S_{s-e}[\Psi,X_i] \,. 
\end{equation}
The combined system and environment is initialized in some initial state, which could even be a factorized state $\rho_s \otimes \rho_e$. This factorized form will fail to hold under evolution, which is generated by the joint system-environment Hamiltonian obtained from the above action. 

If we integrate out the environment degrees of freedom $X_i$,  we will end up with a density operator for the system. This system density operator will itself undergo a non-unitary evolution. The paradigm for such open system dynamics was laid out in~\cite{Feynman:1963fq}. A key observation in this work was that the natural way to describe the evolution of the system is in terms of a Schwinger--Keldysh path integral, with a non-trivial interaction between the L and R degrees of freedom, which is referred to as the \emph{influence function}. 
\begin{equation}
\begin{split}
&
    \int [D\Psi_R] [D\Psi_L]\, \int [DX_{i,R}] [DX_{i,L}] \, 
    \e^{i\, S[\Psi_R, X_{i,R}]-i\,  S[\Psi_L, X_{i,L}]}\\
&\quad 
= 
    \int [D\Psi_R] [D\Psi_L]\,\e^{i\, S[\Psi_R]-i\,  S[\Psi_L] + i\, S_{IF}[\Psi_R,\Psi_L]}\,.
\end{split}
\end{equation}
The influence functional is clearly induced onto the system owing to the coupling with the environment.

This paradigm has broad relevance, ranging from the analysis of environment-induced effects, cosmology, and thermal systems. An approach that is usually taken in these contexts is to assume that the environment's dynamics is suitably Markovian. By this, one means that the system-environment interactions are such that any information exchanged between them is rapidly dissipated or ``forgotten'' by the environment. In other words, the environment has no long-term memory. This implies that the system density matrix will evolve by a trace-preserving positive map (a quantum channel). In quantum mechanics, one can, with these assumptions, write down a master equation for the evolution, generalizing the standard unitary evolution. This takes the form
\begin{equation}
i\,\dv{\rho}{t} = \comt{H,\rho}+ i\, \sum_{a,b} \, \Gamma_{ab}\left( L_b\,\rho\, L_a^\dag - \frac{1}{2}\, L_a^\dag \, L_b\,\rho - \frac{1}{2}\, \rho  \, L_a^\dag\, L_b  \right)\,.
\end{equation}
Here, $L_a$ and $L_b$ are positive operators indexed by $a$, $b$, and $\Gamma_{a,b}$ are effective couplings. The evolution can be seen to be trace-preserving.

In general, the Markovianity assumption can fail, especially in situations where the environment has some long-lived modes dominating its late-time dynamics. Integrating these out will clearly lead to some non-local effects. In such situations, the influence functional is more directly useful. 

Note that in the absence of the influence functional, the dynamics of the system is factorized between the ket and bra (R and L). All non-trivial correlations in this case are all captured by the initial density matrix. 
The factorization fails when we have influence functionals. However, these cannot be completely arbitrary, since that would violate the microscopic unitarity (before integrating out the environment). The influence functional is constrained by the largest time equation, which leads to a set of Lindblad conditions in the case of Markovian dynamics; cf.~\cite{Baidya:2017eho} for an analysis in the context of renormalization of an open QFT.  

An arena that is well suited to develop the structural aspects of open effective field theories (open EFTs) is holography~\cite{Jana:2020vyx,Loganayagam:2022zmq}. We imagine the environment to be a large $N$ holographic field theory, which is strongly coupled. A class of such theories has a simple dual description in terms of gravitational dynamics in an asymptotically AdS spacetime. The prototype example of this is the four-dimensional $\mathcal{N} =4$ SYM theory with gauge group $\text{SU}(N)$. In the large $N$ limit, the theory has a planar diagram expansion with 't Hooft coupling $\lambda = g_{YM}^2\, N$. The dual gravitational theory lives on an AdS spacetime with curvature length scale $\ell_\text{AdS}$ which is hierarchically larger than the string and Planck scales, $\ell_s$ and $\ell_P$, respectively. The parameters are related through, 
\begin{equation}
\lambda \sim \left( \frac{\ell_\text{AdS}}{\ell_s} \right)^4\,, \qquad 
N^2 \sim \left( \frac{\ell_\text{AdS}}{\ell_P} \right)^4\,.
\end{equation}
We will describe below the general setup (cf.~Section \ref{sec:grsk}) for computing Schwinger--Keldysh correlators in such holographic field theories. These enter as the effective couplings of the open quantum system coupled to a holographic environment.

\section[Thermal states and advanced/retarded basis]{Thermal states and advanced/retarded basis\\
\normalfont{\textit{Mukund Rangamani}}}\label{sec:PFbasis}
An interesting and important class of examples is the behavior of systems coupled to thermal environments. To understand these, we can specialize to considering the initial state to be a thermal density operator
\begin{equation}
\rho_\beta = \e^{-\beta\, H}\,.
\end{equation}
%

\subsection{The KMS condition}\label{sec:kms}
Since the thermal density operator involves an evolution in the Euclidean time direction (by an amount $\beta$), we can define a useful notion of conjugating operators through the density operator, and inserting them at imaginary time values: 
\begin{equation}\label{eq:kmsslide}
\mathcal{O}(t-i\beta) = \rho_\beta^{-1}\, \mathcal{O}(t)\, \rho_\beta\,.
\end{equation}
If we view the conjugation operation as passing an operator through the density matrix, then the motion is counterclockwise along the contour. In fact, viewing the density operator itself as an evolution in imaginary time, the conjugation pictorially can be  visualized by sliding the smallest-time operator across the initial state:
\begin{equation*}
\includegraphics[width=\textwidth]{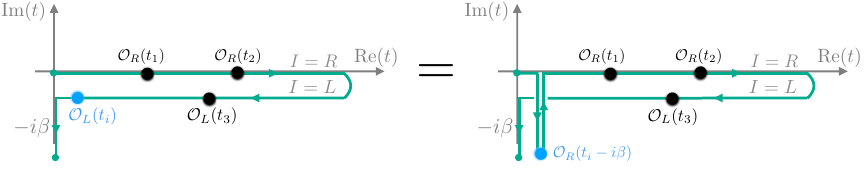}
\end{equation*}

As before, we define the one-sided thermal SK (`1-OTO') path integral as
\begin{equation}
    \mathcal{Z}_\text{SK}^\beta[J_R , J_L] = \Tr( U[J_R] \,\rho_\beta\, (U[J_L])^\dagger )\,.
\end{equation}
This path integral is subject to a KMS condition and unitarity constraints. To describe these, one typically works with the following (advanced/retarded) basis of operators:
\begin{equation}
    \begin{split}
         \mathcal{O}_\adv &\equiv \mathcal{O}_R- \mathcal{O}_L \,,\\
         \mathcal{O}_\ret &\equiv \left(\frac{ \mathcal{O}_R + \mathcal{O}_L}{2} \right) - \frac{1}{2} \coth \left( \frac{i \beta \partial_t}{2} \right) \left( \mathcal{O}_R - \mathcal{O}_L\right) \,.
    \end{split}
\end{equation}
This choice makes manifest the following constraints:
\begin{itemize}[label=$\diamond$]
\item The largest-time equations discussed earlier continue to hold in the form
\begin{equation}
\big\langle {\cal T}_{\cal C} \; {\color{RoyalBlue}\mathcal{O}_{\adv}(t_f)} \, \mathcal{O}_{I_1}(t_1) \cdots \mathcal{O}_{I_n}(t_n) \big\rangle_\beta = 0 \qquad (t_f > t_1, \ldots t_n)\,.
\end{equation}
\item In addition, by virtue of the conjugation property~\eqref{eq:kmsslide} one has another constraint, the thermal smallest time equation, or equivalently the KMS condition 
\begin{equation}
\big\langle {\cal T}_{\cal C} \; {\color{RoyalBlue}\mathcal{O}_\ret(t_i)} \, \mathcal{O}_{I_1}(t_1) \cdots \mathcal{O}_{I_n}(t_n)   \big\rangle_\beta = 0 \qquad (t_i < t_1, \ldots t_n)\,.
\end{equation}
This follows from $ \rho_\beta\, \mathcal{O}(t_i)= \mathcal{O}(t_i-i\beta) \rho_\beta = \e^{-i\beta\partial_{t_i}} \mathcal{O}(t_i) \rho_\beta$.
\end{itemize}

For illustration, consider two-point functions. The matrix of all possible SK two-point functions in this basis is
\begin{equation}
\begin{split}
\mathbb{G}_\beta(t,t')&\equiv 
    \begin{pmatrix}
        \langle {\cal T}_{\cal C} \, \mathcal{O}_\ret(t) \mathcal{O}_\ret(t') \rangle 
        &
        \langle {\cal T}_{\cal C} \, \mathcal{O}_\ret(t) \mathcal{O}_\adv(t') \rangle 
        \\
        \langle {\cal T}_{\cal C} \, \mathcal{O}_\adv(t) \mathcal{O}_\ret(t') \rangle
        &
        \langle {\cal T}_{\cal C} \, \mathcal{O}_\adv(t) \mathcal{O}_\adv(t') \rangle
    \end{pmatrix}
    \\&= 
    \begin{pmatrix}
        0 
        &
        iG^R(t-t')
        \\
        iG^A(t-t')
        &
        0
    \end{pmatrix}\,,
    \end{split}
\end{equation}
where
\begin{equation}
\begin{split}
iG^R(t) 
&= 
    \Theta(t) \langle [\mathcal{O}(t), \mathcal{O}(0)] \rangle_\beta
    \\
iG^A(t) 
&= 
    - \Theta(-t) \langle [\mathcal{O}(t), \mathcal{O}(0)] \rangle_\beta\,.
\end{split}
\end{equation}

We should note that the KMS sliding rule directly leads to the fluctuation-dissipation relation, for 
\begin{equation}
\begin{split}
    \Tr (\rho_\beta\, B(0)\, A(t)) &= \Tr(\rho_\beta\, A(t-i\,\beta) B(0)) \\
\Longrightarrow \;\;
    \expval{\anti{A(\omega_1),B(\omega_2)}}_\beta &= \coth(\frac{\beta\,\omega_1}{2})\, \expval{\comt{A(\omega_1),B(\omega_2)}}_\beta\,.
\end{split}
\end{equation}

At higher points the fluctuation dissipation relations will mix the correlators from across OTO contours differing by one timefold. Generically, all the $n!$ correlators can be broken up into KMS orbits of length $n$. The number of independent thermal $n$-point function is therefore $(n-1)!$. It is rather non-trivial combinatorial problem to decompose $(n-1)!$ into equivalence classes under both unitarity and KMS rules, analogous to \eqref{eq:OTOclassification}. This was done in \cite{Haehl:2017eob}.

To give just one example, we noticed that the set of three-point Wightman functions involved both 1-OTO and 2-OTO observables. However, all the 2-OTO correlators can be related in this case to 1-OTO correlators using the KMS condition. The first independent set of correlators whose KMS orbits are confined to 2-OTO contours occur for four-point functions. The chaos correlator, which we discuss below, is the paradigmatic example of this. 

\subsection{One-sided thermal basis}

Define the one-sided thermal $k$-OTO path integral as
\begin{equation}
    {\cal Z}_\text{one-sided}^\beta[J_{1}, \ldots , J_{2k}] = \text{Tr} \left[U^\dagger[J_{2k}] U[J_{2k-1}] \cdots U^\dagger[J_{2}] U[J_{1}] \,\rho_\beta\, \right]\,.
\end{equation}
This path integral is subject to a single KMS condition and $2k-1$ unitarity conditions. To describe these, we work with the following generalized retarded/advanced basis:
\begin{equation}
    \begin{split}
         \mathcal{O}_{\adv(\ell)} &\equiv \mathcal{O}_{\ell+1}- \mathcal{O}_{\ell} \qquad (\ell = 1 , \ldots , 2k-1) \,,\\
         \mathcal{O}_\ret &\equiv \left(\frac{ \mathcal{O}_{2k} + \mathcal{O}_{1}}{2}\right) - \frac{1}{2} \coth \left( \frac{i \beta \partial_t}{2} \right) \left( \mathcal{O}_{2k} - \mathcal{O}_{1}\right) \,,
    \end{split}
\end{equation}
We refer to $\mathcal{O}_{\adv(\ell)}$ as the set of $2k-1$ {\it advanced} combinations and $\mathcal{O}_r$ as the {\it retarded} combination. This basis again diagonalizes the unitarity and KMS condition for the one-sided path integral; that is, we have the following constraints:
\begin{itemize}[label=$\diamond$]
    \item There are $k$ largest-time equations of the form
    \begin{equation}
        \big\langle {\cal T}_{\cal C} \; {\color{RoyalBlue}\mathcal{O}_{\adv(\ell=2m-1)}(t_f)} \, \mathcal{O}_{I_1}(t_1) \cdots \mathcal{O}_{I_n}(t_n) \big\rangle_\beta = 0 \qquad (t_f > t_1, \ldots t_n)\,,
    \end{equation}
    where $m=1,\ldots,k$ and $\mathcal{O}_{I_i}$ are arbitrary operator combinations on the contours, and ${\cal T}_{\cal C}$ denotes $k$-OTO contour ordering. These constraints follow from unitarity: $U^\dagger[J_{\ell+1}]U[J_{\ell}] = 1$ when $J_{\ell+1}=J_\ell$ since $\mathcal{O}_\adv^{(\ell)}$ is sourced by $\frac{1}{2}(J_{\ell+1}+J_\ell)$. Note that, strictly speaking, $\mathcal{O}_{\adv(\ell)}(t_f)$ does not need to be to the future of {\it all} other operators, but only to the future of those operators that have a leg on contour segments $\ell$ or $\ell+1$. In other words, the largest time equation can be read as a `sliding rule', which moves operators around a {\it future turning point} of the contour. For example:
    \end{itemize}
    \begin{equation*}
\includegraphics[width=\textwidth]{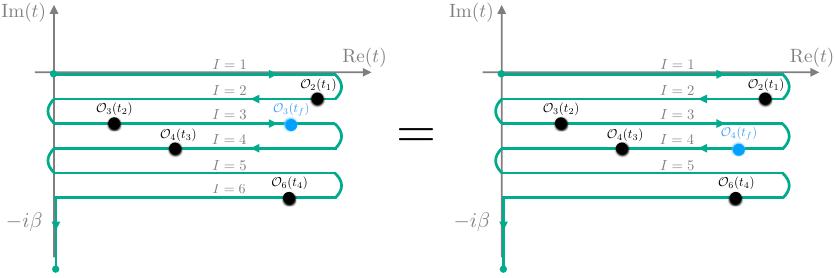}
\end{equation*}
\begin{itemize}[label=$\diamond$]
    \item There are $k-1$ smallest-time equations,
    \begin{equation}
        \big\langle {\cal T}_{\cal C} \; {\color{RoyalBlue}\mathcal{O}_{\adv(\ell=2m)}(t_i)} \, \mathcal{O}_{I_1}(t_1) \cdots \mathcal{O}_{I_n}(t_n)   \big\rangle_\beta = 0 \qquad (t_i < t_1, \ldots t_n)\,,
    \end{equation}
    where $m=1,\ldots, k-1$. These follow again from unitarity as above. Again, $t_i$ only needs to be the smallest time among those insertions that have legs on contour segments $\ell$ or $\ell+1$. Pictorially, this is a `sliding rule' that takes operators around a {\it past turning point} of the contour.
    \item There is one KMS condition,
    \begin{equation}
        \big\langle {\cal T}_{\cal C} \; {\color{RoyalBlue}\mathcal{O}_{\ret}(t_i)} \, \mathcal{O}_{I_1}(t_1) \cdots \mathcal{O}_{I_n}(t_n)   \big\rangle_\beta = 0 \qquad (t_i < t_1, \ldots t_n)\,.
    \end{equation}
    This follows the same way as in the 1-OTO case. Inside a $k$-OTO path ordered expectation value, we can also write the relevant condition as: $\mathcal{O}_{2k}(t_i) = \mathcal{O}_1(t_i-i\beta)$, assuming $t_i$ is smaller than any other time of operator insertions with legs on contour segments $1$ or $2k$.
\end{itemize}

\subsection{Spectral representation}\label{sec:spectral}

There is a natural basis for thermal correlators that takes into account the KMS constraints and expresses the correlators in each KMS orbit in terms of a spectral function. From our earlier discussion, this should imply that there is a single spectral function for two- and three-point functions, and a pair of spectral functions for four-point functions. The construction is achieved by extracting the statistical factors (Bose-Einstein/Fermi-Dirac) from the correlators. This construction is explained in~\cite{Chaudhuri:2018ymp} for generic $k$-OTOs. We will compile some essential features that we will use in our discussion later. 

Since thermal two-point functions are fixed by a single spectral function, introduce 
\begin{equation}
\int\, \frac{\d^\D p}{(2\pi)^\D}\, \varrho(p)\, \e^{i\,p\cdot(x_1 - x_2)} = 
\expval{\comt{\mathcal{O}(x_1),\mathcal{O}(x_2)}}_\beta\,.
\end{equation}
Using this we can represent the contour-ordered two-point functions as 
\begin{equation}
\begin{split}
\expval{\mathcal{T}_{\cal C}\,\mathcal{O}(x_1) \,\mathcal{O}(x_2) }
&= 
    \Theta_{12}\, M(x_1,x_2) + \text{permutations} \,,
\end{split}
\end{equation}
where $M(x_1,x_2)$ is the matrix of Wightman correlators. It can be reconstructed from the spectral function as 
\begin{equation}
M(x_1,x_2) = \int \frac{\d^\D p_1 \d^\D p_2}{(2\pi)^{2\D}}
\, \varrho(p_1 -p_2)\, \mqty(-1\\-1) \, \e^{i\,p_1\cdot x_1} \otimes \mqty(n_B(\omega_2)\\ 1+ n_B(\omega_2)) \, \e^{i\,p_2\cdot x_2} \,,
\end{equation}
where $n_B$ is the Bose-Einstein statistical factor 
\begin{equation}
n_B(\omega) = \frac{1}{\e^{\beta\omega} -1 } \,.
\end{equation}
Similar expressions can be obtained at higher points. For three-point functions, the spectral function is 
\begin{equation}
\int\, \left[ \prod_{i=1}^3\, \frac{\d^\D p_i}{(2\pi)^\D} \right] 
\varrho(p_1,p_2,p_3)\,  \e^{i\,\sum_{i=1}^3\, p_i\cdot x_i} = 
\expval{\comt{\comt{\mathcal{O}(x_1),\mathcal{O}(x_2)},\mathcal{O}(x_3)} 
}_\beta\,.
\end{equation}
For four-point functions we have 
\begin{equation}
\begin{split}
\int\, \left[ \prod_{i=1}^4\,\e^{i\,\, p_i\cdot x_i} \frac{\d^\D p_i}{(2\pi)^\D} \right]
\varrho_1(p_1,p_2,p_{3},p_4)\,
&= 
    \expval{\comt{\comt{\comt{\mathcal{O}(x_1),\mathcal{O}(x_2)},\mathcal{O}(x_3)},\mathcal{O}(x_4)} }_\beta\,,\\
\int\, \left[ \prod_{i=1}^4\, \e^{i\,\, p_i\cdot x_i}\frac{\d^\D p_i}{(2\pi)^\D} \right]
\varrho_2(p_1,p_2,p_{3},p_4)\, 
&= 
    \expval{\comt{\mathcal{O}(x_1),\mathcal{O}(x_2)}\, \comt{\mathcal{O}(x_3),\mathcal{O}(x_4)} }_\beta\,.\\
\end{split}
\end{equation}

Finally, we introduce another basis closely related to the advanced/retarded basis mentioned earlier, but with the salubrious feature of diagonalizing the KMS constraints. We define the future and past combinations of sources to be 
\begin{equation}
J_F = -((1+n_B)\, J_R - n_B\, J_L) \,, \qquad J_P = -n_B\,(J_R - J_L) \,.
\end{equation}
The operators can likewise be rotated in a similar fashion.

\section[Schwinger--Keldysh contours in gravity: grSK]{Schwinger--Keldysh contours in gravity: grSK\\
\normalfont{\textit{Mukund Rangamani}}}\label{sec:grsk}

We now turn to the question of how to implement the Schwinger--Keldysh path integral contour in holographic settings, as indicated in our discussion of the open QFT paradigm. For this discussion, we will focus primarily on the thermal state of the holographic theory. Furthermore, we will focus exclusively on the Schwinger-Keldysh contour, for which there is now a well understood gravitational prescription. 

We first recall that the thermal state can be prepared by a Euclidean path integral, where the CFT is placed on a geometry with a compact time coordinate. As in our previous discussion, one opens up the contour around $t_E =0$ to allow for the real-time timefolded evolution. 

The Schwinger--Keldysh contour of the field theory maps in the gravitational setting to a boundary condition for the semiclassical quantum gravity path integral. The natural question involves identifying the corresponding bulk geometry, which we recall has to be a saddle point of the gravitational path integral. We will follow the prescription of~\cite{Glorioso:2018mmw}, as explained in~\cite{Jana:2020vyx}. We should note important previous work on the subject~\cite{Son:2002sd,Skenderis:2008dg} and in particular~\cite{vanRees:2009rw}.

\begin{figure}[ht]
\centering
\includegraphics[width=.45\textwidth]{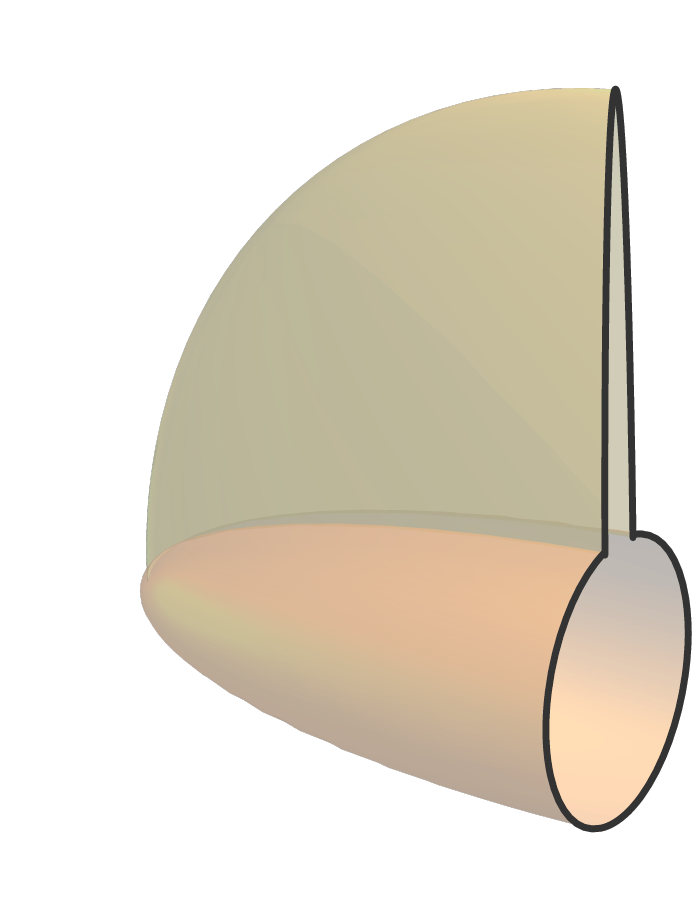}
\hspace{1cm}
\includegraphics[width=.39\textwidth]{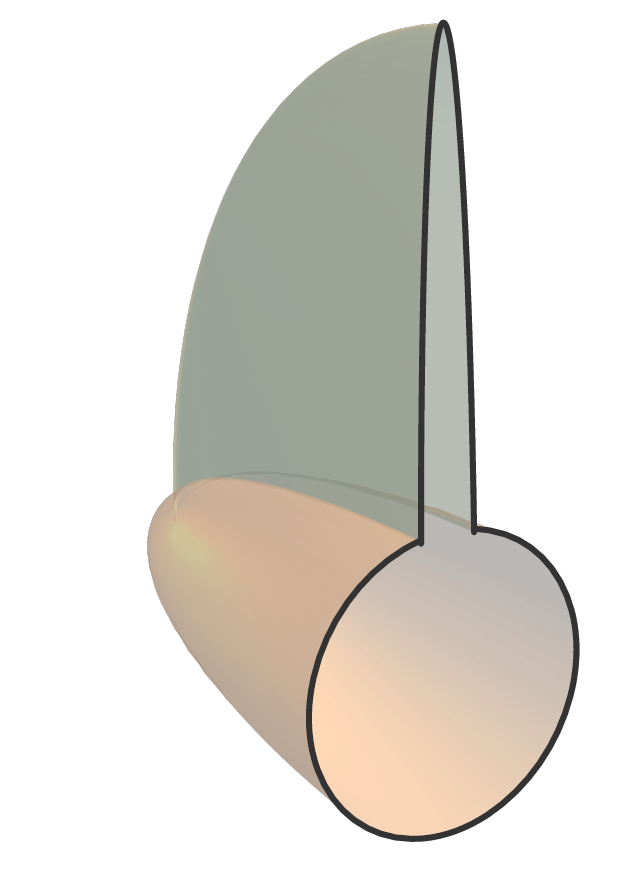}
\includegraphics[width=0.42\textwidth]{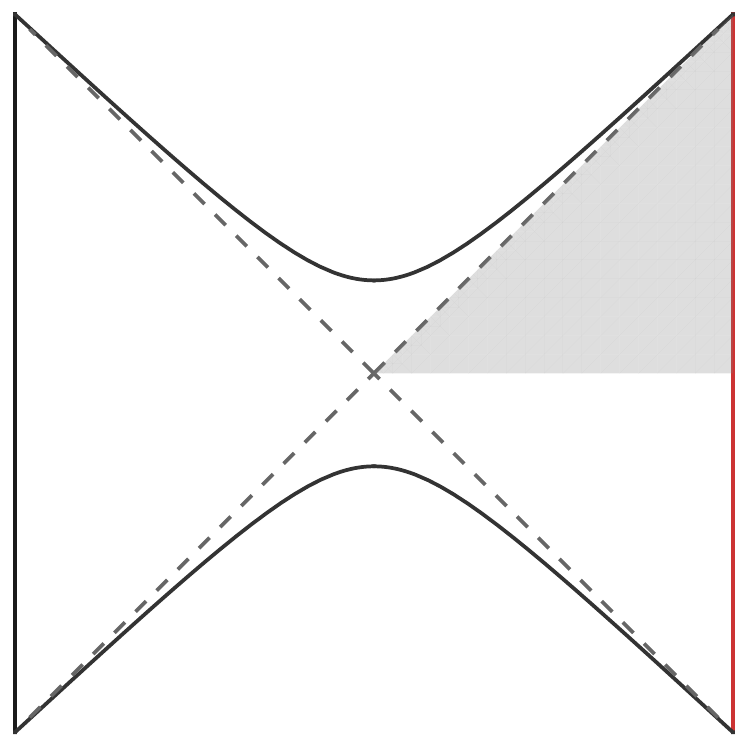}
\hspace{1cm}
\includegraphics[width=0.42\textwidth]{figures/sadspd.pdf}
\begin{picture}(0,0)
\setlength{\unitlength}{1cm}
\put (-8.3,8.5) {$\circlearrowleft\scriptstyle{t_E}$}
\put (-1.6,8) {$\circlearrowleft\scriptscriptstyle{t_E}$}
\put (-8.05,12) {$\uparrow \; t$}
\put (-8.65,11) {$t\; \downarrow$}
\put (-10,12) {L}
\put (-7.8,11) {R}
\put (-2.8,11) {L}
\put (-1.4,13) {R}
\put (-1.63,11) {$\uparrow\; t$}
\put (-2.43,12) {$t\;\downarrow$}
\put (-9.5,4) {$\searrow$ L}
\put (-2,4) {$\nwarrow$ R}
\end{picture}
\caption{The two-sheeted complex grSK geometry shown from two different perspectives. On the top left we display the boundary thermal SK contour, which is filled in the Euclidean portion by the Euclidean black hole geometry (the cigar) and in the Lorentzian section by two copies of the domain of outer communication of the Lorentzian black hole spacetime. The top right panel displays the bulk perspective to emphasize the smooth join of the two sheets of the Lorentzian section. On the bottom panel, we illustrate the Lorentzian sections of the geometry on the Schwarzschild-AdS$_{\D+1}$ Penrose diagram. with the regions pertaining to the $L$ and $R$ sheets of the grSK spacetime shaded. Figure adapted from~\cite{Jana:2020vyx}.}
\label{fig:grsk}
\end{figure}

The thermal state preparation is achieved by a non-trivial saddle of the gravitational path integral. The geometry is the Euclidean black hole spacetime, with the time circle fixed (in an appropriate conformal frame) at the AdS boundary. This circle shrinks to zero in the interior of the spacetime, with the locus of zero proper size corresponding to the bifurcation surface of the event horizon in the Lorentzian continuation. The spacetime has the topology of a disc (often referred to as the cigar geometry). We open up this geometry in the vicinity of $t_E =0$ and patch in two copies of the Lorentzian spacetime, which are glued together along their respective future horizons, see~Fig.~\ref{fig:grsk}.

It is useful to write an explicit set of coordinates and present this solution as a complex two-sheeted geometry. To do so, consider the metric in ingoing Eddington-Finkelstein coordinates
\begin{equation}\label{eq:efads}
\d s^2 = -r^2\, f(r) \, \d v^2 + 2\, \d v\, \d r + r^2\, \d{\bf x}^2 \,, \qquad f(r) = 1 - \frac{r_+^\D}{r^\D} \,.
\end{equation}	
The coordinate $v$ is identified with the time coordinate $t$ on the boundary of the spacetime $r \to \infty$, and follows the Schwinger--Keldysh contour. The radial coordinate in the bulk will be complexified, and we will give a contour prescription to choose a codimension-1 slice. 
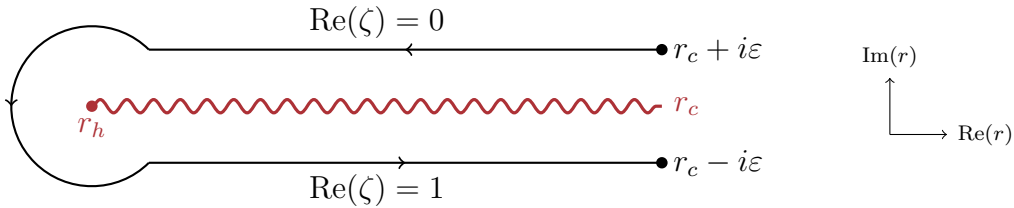
\begin{figure}[t!]
\begin{center}
\begin{tikzpicture}[scale=0.75]
\draw[thick,color=Maroon,fill=Maroon] (-5,0) circle (0.45ex);
\draw[thick,color=black,fill=black] (5,1) circle (0.45ex);
\draw[thick,color=black,fill=black] (5,-1) circle (0.45ex);
\draw[very thick,snake it, color=Maroon] (-5,0) node [below] {$r_h$} -- (5,0) node [right] {$r_c$};
\draw[thick,color=black, ->-] (5,1)  node [right] {$r_c+i\varepsilon$} -- (0,1) node [above] {$\Re(\zeta) =0$} -- (-4,1);
\draw[thick,color=black,->-] (-4,-1) -- (0,-1) node [below] {$\Re(\zeta) =1$} -- (5,-1) node [right] {$r_c-i\varepsilon$};
\draw[thick,color=black,->-] (-4,1) arc (45:315:1.414);
\draw[thin, color=black,  ->] (9,-0.5) -- (9,0.5) node [above] {$\scriptstyle{\Im(r)}$};
\draw[thin, color=black,  ->] (9,-0.5) -- (10,-0.5) node [right] {$\scriptstyle{\Re(r)}$};  
\end{tikzpicture}
\caption{ The complex $r$ plane with the locations of the two boundaries and the horizon marked. The grSK contour is a codimension-1 surface in this plane (drawn at fixed $v$). The direction of the contour is indicated counterclockwise, circling the branch point at the horizon.}
\label{fig:mockt}
\end{center}
\end{figure}

Operationally, we, in fact, upgrade the radial tortoise coordinate to a complex variable, which we refer to as the \emph{mock tortoise} coordinate, $\zeta$. We define this coordinate by the differential relation:
\begin{equation}\label{eq:ctordef}
\frac{\d r}{\d\zeta} = \frac{i\,\beta}{2} \, r^2\, f(r) \,,
\end{equation}	
where $\beta = \frac{4\pi}{\D \,r_+}$ is the inverse temperature of the black hole. While we have indicated the explicit expressions for the case of a Schwarzschild-AdS$_{\D+1}$ black hole, the construction straightforwardly generalizes to any non-extremal black hole, see~\cite{Loganayagam:2022zmq}.

By convention, we choose one of the sheets to have a vanishing real part and the other to have a unit real part. This fixes the normalization used to define $\zeta$. 
\begin{equation}\label{eq:ctorbc}
\lim_{r\to\infty}\zeta(r+i\,\varepsilon) = 0 \,, \qquad \lim_{r\to\infty}\zeta(r-i\,\varepsilon) = 1\,.
\end{equation}	
A section of geometry in the mock tortoise complex plane is illustrated in Fig.~\ref{fig:mockt}. The complexified metric defining the grSK geometry is 
\begin{equation}\label{eq:sadsct}
\d s^2 = -r^2\, f(r) \, \d v^2 +  i\, \beta\, r^2 \, f(r)\,  \d v\, \d\zeta + r^2\, \d{\bf x}^2 \,, \qquad f(r) = 1 - \frac{r_+^\D}{r^\D}\,.
\end{equation}	

One can now give a simple algorithm for computing real-time observables in the holographic CFT using Witten diagrams on the grSK geometry. 
\begin{itemize}[label=$\diamond$]
\item  To begin with one places the gravitational theory on the grSK contour, writing for the bulk action a contour integral
\begin{equation}\label{eq:bilk}
S_\text{bulk} = \oint \d\zeta \int \d^\D x \, \sqrt{-g} \; \mathcal{L}[g_{AB}, \Phi]\,.
\end{equation}	
Here, $\Phi$ denotes the set of bulk fields associated to the CFT operator $\mathcal{O}$, and $g_{AB}$ the bulk metric is dual to the CFT stress tensor. 

\item The main task is to determine the propagators for the linearized fluctuations of the fields $\Phi$ (and metric) around the grSK saddle. To this end, consider the linear wave equation in the original black hole spacetime in ingoing coordinates~\eqref{eq:efads}. Working in Fourier domain with frequencies and momenta denoted $\omega$ and $\vb{k}$, one solves the wave equation. The boundary conditions to be used are ingoing boundary conditions at the horizon (which ensure regularity) and requiring the field to asymptote to an appropriate source at the conformal boundary. We will refer to this as the ingoing boundary to bulk propagator $G_\text{in}$ 
\begin{equation}\label{eq:}
\lim_{r\to \infty} G_\text{in} =1 \,, \qquad \dv{G_\text{in}}{r} \bigg|_{r_+}  =0 \,.
\end{equation}	
\item The ingoing boundary to bulk propagator, being regular at the horizon, does not see the branch cut inherent in the definition of $\zeta$. Therefore, one can immediately write down the ingoing grSK propagator $G_\text{in}(\omega,\vb{k},\zeta)$ with suitable replacements. 
\item The second propagator that we need is the outgoing propagator. To this end we exploit the time-reversal involution of the grSK geometry and deduces that it is given by frequency reversing the ingoing propagator and multiplying with a radial Boltzmann factor
\begin{equation}
G_\text{out}(\omega, \vb{k},\zeta) \equiv G_\text{in}(-\omega, \vb{k},\zeta)\,  \e^{-\beta\,\omega\,\zeta}  \,.
\end{equation}
\item The solution to the field $\Phi$ with L and R Schwinger--Keldysh sources on the respective boundaries then can be written solely in terms of the ingoing propagator. We have  
\begin{equation}\label{eq:phiLR}
\begin{split}
\Phi(\zeta,\omega,\vb{k}) 
&= 
    G_\text{in}(\omega,\vb{k},\zeta) \,  \bigg( (1+n_\omega)\, J_{R} -n_\omega \, J_{L} \bigg)\\& \qquad - G_\text{in}(-\omega,\vb{k},\zeta) \, \e^{\beta\, \omega\, (1-\zeta)}\,n_\omega \bigg( J_{R} - J_{L} \bigg) \,, \\ 
&= 
 - G_\text{in}(\zeta,\omega,\vb{k})\, J_F + G_\text{in}(-\omega,\vb{k},\zeta)\, \e^{\beta\omega\, (1-\zeta)}\, J_P\,.
\end{split}
\end{equation}
\item The final ingredient is the bulk-bulk propagator, which can be obtained using the 
variation of parameters trick. One takes advantage of the solutions of the homogeneous wave equation to construct the solutions to the inhomogeneous one.  This can be efficiently expressed in terms of a linear combination of boundary-bulk Green's functions that are normalizable at one or the other boundary of the grSK geometry. Denoting these as $G_R(\zeta,\omega, \vb{k})$ and $G_L(\zeta,k)$, respectively, we have (setting $k = (\omega, \vb{k})$ for simplicity)
\begin{equation}\label{eq:GLRdef}
\begin{split}
G_R(\zeta,k)
&=
        \e^{\beta\,\omega}\, n_B(\omega) \left(G_\text{in}(\zeta,k) - G_\text{out}(\zeta,k)\right) ,
    \\ 
G_L(\zeta,k)
&=
     -n_B(\omega) \left(G_\text{in}(\zeta,k) - \e^{\beta\omega}\, G_\text{out}(\zeta,k) \right) .
\end{split}
\end{equation}  
The function $G_R$ has a source on the right boundary ($\zeta =1$), while $G_L$ has a source on the left boundary ($\zeta =0$), and they are respectively normalizable at the other end viz.,
\begin{equation}\label{eq:GLRbcs}
\lim_{\zeta \to 0}\, \{G_R, G_L\} = \{0,1\}  \,, \qquad 
\lim_{\zeta \to 1}\, \{G_R, G_L\} = \{1,0\}  \,.
\end{equation}  

The bulk-bulk propagator can be seen to be given by  
\begin{equation}\label{eq:Gbulk}
\begin{split}
G_\text{bulk}(\zeta,\zeta';k)
&=  
   \mathcal{N}(k)\;  \e^{\beta\omega\,\zeta'} 
   \bigg[\Theta(\zeta - \zeta') \, G_L(\zeta, k) \, G_R(\zeta', k) 
   \\& \qquad\qquad\qquad+\Theta(\zeta'- \zeta) \, G_L(\zeta', k)\, G_R(\zeta, k)  \bigg] \,.
\end{split}
\end{equation} 
Here $\Theta(\zeta-\zeta')$ is a contour-ordered step function along the contour depicted in Fig.~\ref{fig:mockt}. The prefactor can be obtained from the Wronskian of the two linearly independent solutions to the homogeneous equation, which we have taken to be the left- and right-normalizable boundary-bulk propagators.
\end{itemize}

Armed with these propagators, one simply writes down Witten diagrams for the boundary correlators with the prescribed sources. Each diagram will have a set of ingoing and outgoing propagators and bulk-to-bulk propagators sewn together at the vertices. This defines the integrand of the Witten diagram, and the integration is over the grSK contour sketched in~Fig.~\ref{fig:mockt}. As demonstrated in~\cite{Jana:2020vyx,Loganayagam:2022zmq,Loganayagam:2024mnj} each such diagram can be expressed as an integral over a single copy of the black hole exterior, along the real contour running from $r_+$ to $\infty$ of an integrand, which is a sum of (multiple) discontinuities of the contour integrand.\footnote{ This is assuming that the vertex functions are simple -- in certain cases there are also localized contributions at the horizons which originate from derivative interactions. See the aforementioned references for more details.}

\section[Applications of \texorpdfstring{$k$}{k}-OTO path integrals]{Applications of \texorpdfstring{$k$}{k}-OTO path integrals\\
\normalfont{\textit{Felix Haehl}}}

We will now discuss some concrete applications of $k$-OTO path integrals, focusing in particular on $k\geq 2$. The goal is to understand some related physical phenomena, which are often of an information-theoretic nature.

\subsection{OTOCs in thermal states}

We begin with a review of out-of-time-order correlation functions and their relation to quantum chaos. The basic ideas were first discussed in \cite{Shenker:2013pqa,Shenker:2013yza,Roberts:2014isa,Maldacena:2015waa}.

\subsubsection{Quantum butterfly effect}

We would like to consider the two states
\begin{equation}
\begin{split}
   \hspace{-1.5cm}  \qquad\qquad\qquad|\Psi \rangle &\equiv V(-T,0) W(0,x) |\beta \rangle \equiv \e^{-iHT}V(0,0) \e^{iHT} W(0,x)  |\beta\rangle = \adjustbox{valign=c}{\includegraphics[width=.1\textwidth]{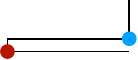}}\,,\\
    |\Phi \rangle &\equiv W(0,x) V(-T,0) |\beta \rangle  \equiv  W(0,x) \e^{-iHT} V(0,0) \e^{iHT}|\beta\rangle = \adjustbox{valign=c}{\includegraphics[width=.1\textwidth]{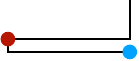}}\,,
\end{split}
\end{equation}
where $|\beta \rangle =  \e^{-\beta H/2}|0\rangle$ is the square root of the thermal density matrix such that $\langle \beta | (\cdots) |\beta \rangle \equiv \text{Tr}[(\cdots) \rho_\beta]$.
We expect that the two states are generically very similar when the time separation $T \ll \beta$. However, if $H$ is a suitably {\it chaotic} Hamiltonian, then the overlap becomes small for large time separations. By the overlap, we mean the following 2-OTO correlation function:
\begin{equation}
\begin{split}
   \hspace{-1.5cm}  \qquad\qquad\qquad\langle \Phi | \Psi \rangle &= \text{Tr} \left[V^\dagger(-T,0)W^\dagger(0,x)V(-T,0)W(0,x)\, \rho_\beta \right]=  \,\adjustbox{valign=c}{\includegraphics[width=.1\textwidth]{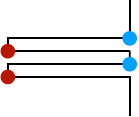}}\\
    &\sim \langle W^\dagger W \rangle_\beta\, \langle V^\dagger V \rangle_\beta \underbrace{\left[ 1 - \frac{c_0}{N} \, \e^{\kappa(v) T} + {\cal O}(N^{-2}) \right] }_{\rightarrow \, 0 \;\, (T \rightarrow \infty)}
    \end{split}
\end{equation}
where $\kappa$ is the Lyapunov exponent. In general, the Lyapunov exponent depends on the spatial separation between $V$ and $W$. Specifically, it is expected to be a function of the `velocity' $v = x / T$, or equivalently of the relative spatial momentum. We will suppress this spatial dependence for now and get back to it later.

Let us now be more precise about the meaning of the above. We canonically purify the system by doubling the Hilbert space ${\cal H} \rightarrow {\cal H}_L \otimes {\cal H}_R$. We use a convention where the time in the $R$-Hilbert space is aligned with global time, $t_{R} = t$, and the time in the $L$-Hilbert space runs in the opposite direction, $t_L = -t$. The purification of the thermal density matrix corresponds to a state $|\text{TFD}\rangle \in {\cal H}_L \otimes {\cal H}_R$. Let us discuss three closely related correlation functions in the doubled system (see, for example, \cite{Shenker:2014cwa}), which make different features of our discussion manifest (square brackets).

\subparagraph{1.\ One-sided correlator [OTO path integral]:} The states we compare can be chosen in various ways; a conveninent choice is to represent all operators in the $R$-Hilbert space:
\begin{equation}
    |\Psi_1 \rangle \equiv  V_R(-T) W_R  (0) |\text{TFD}\rangle \,,\qquad
    |\Psi_2 \rangle \equiv W_R(0) V_R(-T) |\text{TFD}\rangle \,,
\end{equation}
such that their overlap reads as
\begin{equation}
    \langle \Psi_2 | \Psi_1 \rangle 
    = \langle \text{TFD} | V_R^\dagger W_R^\dagger V_RW_R  \, | \text{TFD} \rangle
    = \text{Tr} \left[ V^\dagger W^\dagger VW \rho_\beta \right] \,.
\end{equation}
Written in this way, the overlap $\langle \Psi_2 | \Psi_1 \rangle $ is a thermal 4-point 2-OTO correlation function of the type discussed in previous sections. It compares the effect of acting with $W(0)$ first followed by $V(T)$, versus acting in the opposite order. In a large $N$ system and for `simple' operators $V$, $W$, the difference between these two states is initially negligible, but can become large as $T$ grows. This is the basic feature of the quantum butterfly effect.

\subparagraph{2. Overlap of `in' and `out' states [scattering experiment]:} The one-sided OTOC is closely related to the overlap of an `in' state on the time slice at an early global time and an `out' state at late time:\footnote{Recall $t_L = -t$, which means $V_L(t_{L}=T)$ corresponds to an insertion at global time $-T$. Contour pictures are drawn in the complex `global' $t$-plane.}
\begin{equation}
\begin{split}
   \qquad\qquad\qquad\qquad\qquad |\Psi_{\text{in}} \rangle \equiv W_L(0) V_R  (-T)|\text{TFD}\rangle
    &= \!\!\!\!\!\!\!\!\vcenter{\includegraphics[width=.1\textwidth]{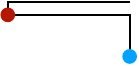}}\\
    |\Psi_{\text{out}} \rangle \equiv  V_L(T) W_R (0) |\text{TFD}\rangle 
&=\!\!\!\!\!\!\!\!\vcenter{\includegraphics[width=.1\textwidth]{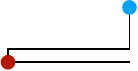}}
\end{split}
\end{equation}
such that
\begin{equation}
\label{eq:otocInOut}
      \hspace{-1.5cm} \qquad\quad\langle \Psi_{\text{out}} | \Psi_{\text{in}} \rangle 
    = \langle \text{TFD} | V_L^\dagger  W_L W_R^\dagger  V_R   | \text{TFD} \rangle
    = \text{Tr} \left[W V^\dagger  \rho_\beta^{1/2}\, W^\dagger V \rho_\beta^{1/2} \,\right]=\adjustbox{valign=c}{\includegraphics[width=.1\textwidth]{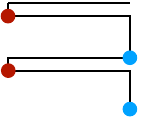}}
\end{equation}
The labels `in' and `out' refer to whether the states are created in the past or future with respect to global time.
This version of the OTOC is most easily interpreted as a scattering problem. The `in' and `out' states are defined on time slices, i.e., they each involve two spacelike separated operator insertions. We can then describe their difference in terms of a 2-to-2 scattering phase; see, e.g., \cite{Stanford:2015owe}.

\subparagraph{3. Two-point function in a perturbed state [probing a shockwave state]:} Finally, consider the state 
\begin{equation}
\label{eq:shockstate}
    |\Phi_V\rangle \equiv V_R(0) |\text{TFD} \rangle \,,
\end{equation}
and write the left-right two-point function of $W$:
\begin{align}
\label{eq:otocProbe2}
    \langle \Phi_V | W_L^\dagger(-T) W_R(T) | \Phi_V \rangle & = \langle \text{TFD} |  W_L^\dagger V_R^\dagger W_R V_R | \text{TFD} \rangle \\&= \text{Tr} \left[ W^\dagger\rho_\beta^{1/2} \, V^\dagger W V \rho_\beta^{1/2} \,\right]\notag
\,.\end{align}
This analytic continuation of the OTOC cannot be written as the overlap of an `in' and `out' state. It should be thought of as a two-point probe correlator evaluated in the state that corresponds to a thermal state perturbation by $V$. 

It is instructive to compute OTOCs using holography. Doing so properly should involve a generalization of the setup described in section \ref{sec:grsk} to geometries with boundary conditions appropriate for the OTO path integral (`grOTO geometry'). This has not yet been studied for the correlators described here. However, the calculation is  easy in a certain approximation. For instance, the state \eqref{eq:shockstate} evolved to a sufficiently late time slice can be approximated by an eternal black hole geometry perturbed by a null shockwave that travels from the right boundary towards the black hole. That is, the gravitational backreaction (graviton exchanges on the grOTO geometry) is effectively reorganized into a gravitational shockwave. The OTOC \eqref{eq:otocProbe2} is then computed by a late-time left-right correlator in an eternal black hole geometry perturbed by a shockwave \cite{Shenker:2013pqa}.

\subsubsection{Six-point OTOCs and shockwave collisions}

Higher-point OTOCs play a natural role when one tries to measure detailed properties of states that are perturbed with respect to some reference state.

A 3-OTO application can be motivated as follows \cite{Haehl:2021prg,Haehl:2021tft,Haehl:2022frr}. Consider the perturbed thermofield double state:
\begin{equation}
    |\Phi_{VW} \rangle \equiv  W_L(-t_{wL}) V_R(-t_{vR})|\text{TFD}\rangle \,.
\end{equation}
This state has an interesting interpretation in holography: for large $t_{vR}$ and $t_{wL}$, it corresponds approximately to the eternal AdS black hole perturbed by two shockwaves traveling from the left and right boundaries, respectively, in almost null directions (see figures below). The two shocks may scatter behind the black hole horizon. We can try to characterize this scattering process and how it affects the post-collision geometry. A natural probe would be a left-right two-point function on a late time slice $t_0 \gg t_{wL}, t_{vR}$:
\begin{equation}
\begin{split}
        \langle \Phi_{VW} | &{\cal O}_L(t_0) {\cal O}_R(t_0) | \Phi_{VW} \rangle \\&= \text{Tr} \left[{\color{RoyalBlue}W^\dagger(t_{wL})}{\color{ForestGreen}{\cal O}(-t_0) }{\color{RoyalBlue}W(t_{wL})}  \, \rho_\beta^{1/2} \,{\color{Maroon}V^\dagger(-t_{vR})}{\color{ForestGreen}{\cal O}(t_0)}{\color{Maroon}V(-t_{vR})} \, \rho_\beta^{1/2} \right]\,,
\end{split}
\end{equation}
where we translated the two-sided correlator into a trace (3-OTO path integral) over a single Hilbert space. On the contour, it looks schematically as follows:
\begin{equation*}
\includegraphics[width=.5\textwidth]{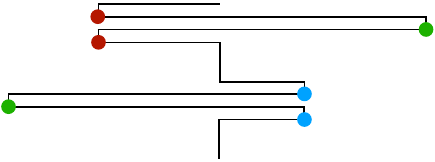}
\end{equation*}
Due to the isometries of the eternal black hole geometry, the result of this calculation should depend on the total separation $\Delta t = t_{wL}+t_{vR}$. One finds that for small $\Delta t$ the six-point function cluster-decomposes into two-point functions to a good approximation. However, when {\it both} $V$ and $W$ each had enough time to scramble, the correlator will deviate significantly from the factorized result. One finds \cite{Haehl:2021tft}:
\begin{equation}
    \hspace{-0.3cm} \langle \Phi_{VW} | {\cal O}_L(t_0) {\cal O}_R(t_0) | \Phi_{VW} \rangle \approx \langle V^\dagger V \rangle_\beta \langle W^\dagger W \rangle_\beta \langle {\cal O} {\cal O} \rangle_\beta 
    \left[ \frac{1}{1 + \alpha\, G^2 \Delta_V \Delta_W \, \e^{\Delta t}}\right]^{2\Delta_{\cal O}}
\end{equation}
with some theory- and dimension-dependent number $\alpha$ and  the Newton constant $G$. Importantly, this result deviates from the product of two-point functions by a large amount after {\it two} scrambling times, i.e., when $\Delta t \sim t_{*W} + t_{*V}$ with $t_{*{\cal O}} \equiv \log(\sqrt{\alpha} G \Delta_{\cal O})$.
This deviation indicates that the gravitational interaction between the two shockwaves is very disruptive. Indeed, the decorrelation of the left-right two-point function can be interpreted as the disappearance of spacetime connecting the two sides. A more detailed gravitational analysis indicates that the total spacetime volume within the causal future of the collision point rapidly decreases to an exponentially small value because of the strong non-linear gravitational backreaction. The two situations can be pictured as follows:\footnote{We are glossing over some details here: initially, as we increase $t_{wR}$ and $t_{wL}$ from zero, the gray spacetime region grows (roughly linearly). After one scrambling time, the gravitational backreaction becomes strong and the singularity ``bends down'' in such that the grey region begins to shrink. After two scrambling times, it has shrunk to an exponentially small value, as shown on the right \cite{Haehl:2022frr}.}
\begin{equation*}
\includegraphics[width=.8\textwidth]{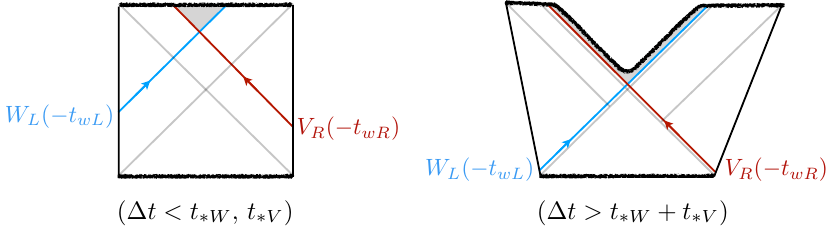}
\end{equation*}

Can we understand this process using a more conventional language appropriate to describe the 2-to-2 scattering process? An immediate obstacle is the ill-definedness of asymptotic out states due to the spacetime singularity. Initially (for small $\Delta t$) this might not be a serious issue because the singularity is far from the collision point; but for large $\Delta t$, the singularity ``bends down'' exponentially close to the collision point. Characterizing the collision in terms of an S-matrix now seems quite problematic. It is no coincidence that the six-point OTOC considered above {\it cannot} be written as the overlap of an in- and an out-state. Relatedly, the four-point OTOC (which {\it can} be written as such an overlap, see \eqref{eq:otocInOut}) is not sensitive to the change of behavior at two scrambling times.
It is an interesting question to investigate how to better understand this process in the language of the S-matrix.

\subsection{The effective theory of chaos and eikonal scattering}

In chaotic systems, it seems to be the case that the SK (or OTO) path integral computing out-of-time order correlators is dominated by few collective modes, which we will refer to as ``scramblons''. We will now define these modes and their effective description, illustrating how their dynamics gives rise to the butterfly effect. It is interesting to note that scramblons live inherently on the real-time contour \cite{Maldacena:2016upp,Stanford:2021bhl,Gu:2021xaj,Choi:2023mab,Gao:2023wun}.

\subsubsection{Generalities}

For concreteness, let us consider the following OTOC:
\begin{equation}
\label{eq:2dOTOC}
    \text{OTOC}(T,x) \equiv \frac{\text{Tr}\left[ V^\dagger(0,0) W^\dagger(T,x) V(0,0) W(T,x) \,\rho_\beta \right]}{\text{Tr}\left[ V^\dagger V \rho_\beta \right]\text{Tr}\left[ W^\dagger W\rho_\beta \right]} \,.
\end{equation}
To compute this, we parameterize the thermal 2-OTO contour using a {\it contour time} coordinate $s \in [0,4T+\beta]$ such that the usual complex time $t$ takes the following values along the contour:
\begin{equation}
 t_I(s) = \begin{sqcases}
     s\,\, \qquad\qquad\qquad &(0\leq s \leq T)\,, \\
  2T-s \qquad\qquad& ( T \leq s \leq 2T)\,, \\
  s-2T \qquad\qquad& (2T \leq s \leq 3T)\,, \\
  4T-s \qquad\qquad& (3T \leq s \leq 4T)\,, \\
  -i(s-4T) \qquad& (4T \leq s \leq 4T+ \beta)\,,
 \end{sqcases}
\end{equation}
where $I=1,\ldots ,5$ labels the five segments of the contour ($I=5$ is Euclidean):
\begin{equation*}
\includegraphics[width=.65\textwidth]{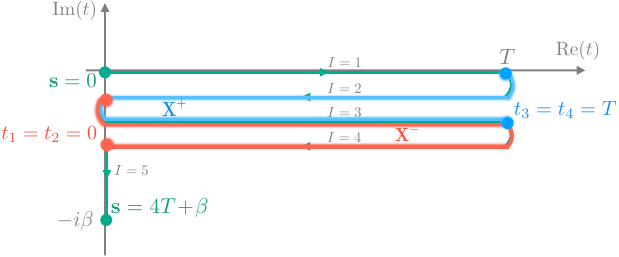}
\end{equation*}
Conceptually, it can be useful to ``unwrap'' this contour, which makes the flow of contour time more manifest:
\begin{equation*}
\includegraphics[width=.8\textwidth]{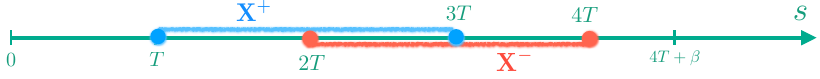}
\end{equation*}

Consider now a (schematic) path integral along this contour:
\begin{equation}
    \hspace{-0.5cm}{\cal Z}[J_I] = \int 
 [D\phi_I] \; \exp \left\{ i S[\phi_1,J_1]{-} i S[\phi_2,J_2] {+} i S[\phi_3,J_3] {-} i S[\phi_4,J_4] {-} S_E[\phi_5,J_5]
 \right\}\,,
\end{equation}
where, for example, $S[\phi_1,J_1] = \int_0^T \d s \int \d^{\D-1}x \, ({\cal L}[\phi_1] + J_1 \phi_1) $, and so on.

In the above path integrals, $\phi_I$ is a stand-in for the various fields we need to integrate over. Consider the saddle point solution $\phi_\star$ relevant for the thermal state. In the rest of this section we will illustrate the following picture:
\begin{itemize}
\item[]{\it Effective description of scrambling:} Assume that the theory has many degrees of freedom and exhibits fast scrambling. Consider the saddle point corresponding to the thermal state. Then the OTO path integral computing \eqref{eq:2dOTOC} is dominated by the exchange of soft mode fluctuations $\delta \phi_I$ defined on the thermal OTO contour, which grow or decay exponentially in time. Their quadratic action becomes parametrically small around the scrambling time $t_* \sim \log N$. 
\end{itemize}
Specifically, the fluctuations described above are of the form
\begin{equation}
 \delta_+ \phi_I(t_I,p) = X^+(p) \, \e^{-\kappa(p) t_I} \,,\qquad \delta_- \phi_I(t_I,p) = X^-(p)\, \e^{\kappa(p)\, (t_I-T)} \,.  
\end{equation}
where $\kappa$ is the quantum Lyapunov exponent and $X^\pm(p)$ parametrize the strength of the fluctuation as a function of spatial momentum. These fluctuations individually have a vanishing quadratic action, but their interaction leads to an effective `eikonal' action that is proportional to $X^+ X^-$ and exponentially small at late times:
\begin{equation}
    iS_{\text{eik}}[X^+,X^-] = N \int \d p \;  c(\kappa)\, \e^{\frac{i\pi\kappa}{2} - \kappa T} \, X^+(p) X^-(-p) \,.
\end{equation}
The exponential suppression $\e^{-\kappa T}$ means that the large $N$ saddle point approximation breaks down for the contribution of these modes to the OTO path integral. We will account for this by performing the path integral over the modes $X^\pm$ fully. The phase factor $\e^{\frac{i\pi\kappa}{2}}$ in the eikonal action is a universal feature when $\kappa \neq 1$ and a manifestation of the KMS condition \cite{Gu:2021xaj,Choi:2023mab}.

\subsubsection{A toy model}

Although we would ultimately like to understand the physics discussed here in theories such as ${\cal N}=4$ SYM (let alone QCD), we will discuss a toy model instead, which exhibits some of the crucial features.

The Sachdev--Ye--Kitaev (SYK) model describes disordered interactions between $N$ Majorana fermions \cite{Sachdev:2015efa,Maldacena:2016hyu}. After disorder averaging, the large $N$ model can be solved in terms of a collective variable $G(\tau_1,\tau_2)$ that is bilocal in time. It gives the mean-field value of the two-point functions averaged over all species of fermions and it satisfies the following Schwinger-Dyson equation:
\begin{equation}
    \hat{G}(\omega_n) = -\frac{1}{i\omega_n + \hat{\Sigma}(\omega_n) } \,,\qquad
    \Sigma(\tau,\tau') 
    = J^2 \, G(\tau,\tau')^{q-1} \,,
\end{equation}
where the self-energy $\Sigma(\tau,\tau') \equiv \sum_n \e^{-i\omega_n (\tau-\tau')} \hat{\Sigma}(\omega_n)$ and $\omega_n = \frac{1}{2} + in$. Pictorially:
\begin{equation*}
\includegraphics[width=.8\textwidth]{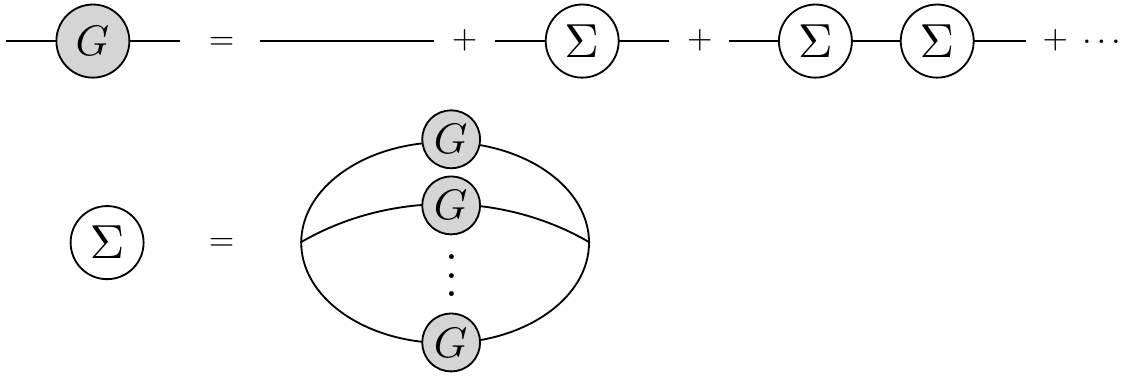}
\end{equation*}

Since $G(\tau_1,\tau_2)$ is the full two-point function, we can use it to build higher-point functions. For example, a four-point turns out to take the following form:
\begin{equation*}
\includegraphics[width=.9\textwidth]{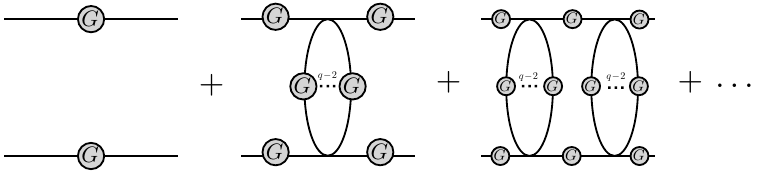}
\end{equation*}

At large $J^2$, the $i\omega_n$ can be neglected and the equations become invariant under time reparametrizations:
\begin{equation}
    \begin{split}
        \tau \; &\mapsto \; f(\tau)\,, \\
        G(\tau,\tau') \;&\mapsto [f'(\tau)f'(\tau')]^{\frac{1}{q}} \, G(f(\tau),f(\tau'))\,,\\
        \Sigma(\tau,\tau') \;&\mapsto [f'(\tau)f'(\tau')]^{1-\frac{1}{q}} \, \Sigma(f(\tau),f(\tau'))\,.
    \end{split}
\end{equation}
It is interesting to perform perturbation theory in $1/(\beta J)$ around the strict conformal limit. The $1/(\beta J)$ fluctuations lift the degeneracy and lead to a small action cost associated with any given reparametrization $f(\tau)$. The action is the simplest expression that preserves an $\text{SL}(2,\mathbb{R})$ symmetry of the conformal saddle:
\begin{equation}
\label{eq:SchwarzianAction}
    S = -C \int \d\tau \; \{ f(\tau),\tau \} \,, \qquad\quad C \propto \frac{N}{\beta J} \,.
\end{equation}
This is vaguely similar to what happens in AdS/CFT: while the bulk theory at small $G_N$ is approximated by classical gravity, there are some fluctuations around classical saddles that are enhanced, namely the massless stringy modes controlled by $\ell_s / \ell_{\text{AdS}}$.

\paragraph*{Scrambling in the Schwarzian theory.}
The Schwarzian theory \eqref{eq:SchwarzianAction} arises in the strong coupling limit of the SYK model, among other nearly conformal $(0+1)$-dimensional chaotic spin systems \cite{Maldacena:2016hyu}. 
The Schwarzian path integral is over time reparametrizations $t \rightarrow f(t)$.

Let us now work on the thermal 2-OTO contour appropriate for computing the OTOC (following \cite{Choi:2023mab}). The saddle point corresponding to the thermal state is:
\begin{equation}
    f_I(s) = \tanh \left( \frac{t_I(s)}{2} \right) \qquad (I=1,\ldots,5)\,. 
\end{equation}
The quadratic action of fluctuations $t_I(s) \rightarrow t_I(s) + \epsilon_I(t_I(s))$ is
\begin{equation}
\label{eq:SquadSchwarzian}
    \begin{split}
    iS_{\text{quad}} &= \frac{iC}{2} \int_0^{4T+\beta} \d s \, \frac{\d t_I}{\d s} \;  \epsilon_I(t_I(s)) \left( \partial_{t_I}^4 - \partial_{t_I}^2\right)  \epsilon_I(t_I(s)) \,.
    \end{split}
\end{equation}
From this action we can read off the real-time propagators of the Schwarzian mode along the contour. The contour-ordered propagators can all be expressed in terms of Schwarzian Green's functions \cite{Haehl:2018izb}:
\begin{equation}
\begin{split}
  iG^R_\epsilon(t) &= \Theta(t) \, \langle [\epsilon(t),\, \epsilon(0)] \rangle_\beta
    = \frac{1}{iC} \, \Theta(t) \, (t- \sinh t)  \sim -\frac{1}{2iC} \, \Theta(t) \, \e^t\,,
    \\
    iG^A_\epsilon(t) &{=} -\Theta(-t) \, \langle [\epsilon(t),\, \epsilon(0)] \rangle_\beta 
    {=} {-}\frac{1}{iC} \,\Theta(-t)\, (t{-} \sinh t)  \sim -\frac{1}{2iC} \, \Theta(-t) \, \e^{-t}\,.
\end{split}
\end{equation}
The Schwarzian mode couples to external operators universally via time reparametrizations. This defines natural {\it bilocal} operators, which replace pairwise operator insertions:
\begin{equation}
    {\cal O}_\Delta(t_I) {\cal O}_\Delta({t}_J) \, \stackrel{f}{\longrightarrow} \, {\cal B}_{IJ}^\Delta (t_I,{t}_J) = \left[ -\frac{f_I'(t_I) f_J'({t}_J)}{(f_I(t_I) - f_J({t}_J))^2} \right]^\Delta \,,
    \label{eq:Bschwarzian}
\end{equation}
which we can linearize as
\begin{equation}
\label{eq:Blinearized}
   \hspace{-0.5cm}
   {\cal B}_{IJ}^\Delta(t_I, {t}_J) 
    {=} \langle {\cal O}_\Delta(t_I) {\cal O}_\Delta({t}_J) \rangle_\beta \left[ 1 {+} \Delta \Big[ \epsilon_I'(t_I) {+} \epsilon_J'({t}_J) {-} \frac{\epsilon_I(t_I) {-} \epsilon_J({t}_J)}{\tanh \left( \tfrac{t_I{-}{t}_J}{2} \right)} \Big]{+}\mathcal{O}(\epsilon^2)\right].
\end{equation}
In the OTOC, we consider these bilocal operators for the case where the two operators are inserted at the same time, only separated by a small imaginary regulator that distinguishes the contours. In the above equations, this means $J=I+2$ and $t_I = t$, $t_J = t-2i\varepsilon$. The linearized bilocal becomes
\begin{equation}
\begin{split}
    {\cal B}^\Delta_{I,I+2}(t,t{-}2i\varepsilon) &= \frac{\left[ 1 {+} 2\Delta \left( \partial_t \epsilon_{\av(I,I+2)}(t) {-}\frac{\epsilon_{\dif(I,I+2)}(t)}{2i\,\sin \varepsilon} \right) {+} \mathcal{O}(\epsilon^2)
    \right]}{(2 \sin\varepsilon)^{2\Delta}}\,,
\end{split}
\end{equation}
where we use a basis of the form
\begin{equation}
    \epsilon_{\av(I,J)}(t) = \frac{\epsilon_I(t) +\epsilon_{J}(t)}{2} \,, \qquad \epsilon_{\dif(I,J)}(t) = \epsilon_I(t) - \epsilon_{J}(t) \,.
\end{equation}
In the $\varepsilon \rightarrow 0$ regularization, the contribution due to the difference operator dominates. We can say that the external operator insertions act as sources for $\epsilon_{\dif(I,I+2)}$. This means, for example, that the leading connected contribution to the OTOC is approximately
\begin{equation}
\label{eq:OTOCschw}
    \text{OTOC}(T)
    = \langle VV \rangle_\beta \langle WW \rangle_\beta \left[ 1 - \frac{\Delta_V \Delta_W}{\sin^2 \varepsilon} \, \left\langle {\cal T}_c\;
    \epsilon_{\dif(1,3)}(T) \, \epsilon_{\dif(2,4)}(0)
    \right\rangle_\beta + \ldots \right] \,.
\end{equation}
Note that despite having two difference operators, they are obstructing each other in such a way along the contour that their two-point function does not vanish. Indeed, one finds:
\begin{equation}
    \left\langle {\cal T}_c\;
    \epsilon_{\dif(1,3)}(T) \, \epsilon_{\dif(2,4)}(0) 
    \right\rangle_\beta = -\langle [\epsilon(T),\epsilon(0)]\rangle_\beta
    \sim \frac{1}{2iC} \, e^T\,.
\end{equation}
for large $T$.\\

Let us explain the above argument in a different way (see \cite{Maldacena:2016upp}). The action \eqref{eq:SquadSchwarzian} has soft modes that depend exponentially on time:\footnote{We choose convenient (but arbitrary) normalizations such that $\delta_+\epsilon \sim X^+$ at time $t_I=0$ and $\delta_- \epsilon \sim - X^-$ and $t_I=T$.}
\begin{equation}
    \delta_+\epsilon_I = X^+ \, \e^{-t_I(s)} \,,\qquad
    \delta_-\epsilon_I = -X^- \, \e^{t_I(s) - T} \,.
\end{equation}
Along each segment $I$, these modes have vanishing action. This is related to the unbroken $\text{SL}(2,\mathbb{R})$ symmetry of the Schwarzian action. However, the way these modes contribute to the path integral is slightly more subtle: we can think of $W(T)$ as creating an `advanced' perturbation $\delta_+\epsilon_I$, which grows towards the past; this excitation is  small, of order $\e^{-T}$. However, once sourced on contour segments $I=2$ and $I=3$, the effect of this mode will be felt by the early-time operator $V(0)$ on segment $I=3$ (see figure above). At $t=0$, the mode $\delta_+ \epsilon_I$ has
grown to ${\cal O}(1)$ and can therefore have a large effect. Similarly, the operator $V(0)$ acts as a source for the `retarded' perturbation $\delta_- \epsilon$, which grows large ${\cal O}(1)$ as it propagates along the contour to $t=T$. At any point along segment $I=3$ the interaction between the two modes is ${\cal O}(1)$. We summarize this configuration of scramblon modes appropriate for the OTOC computation as follows:
\begin{equation}
\label{eq:epsotoc}
\delta\epsilon^\text{otoc}_I = (\chi_{I2} + \chi_{I3})\, \delta_+ \epsilon_I + (\chi_{I3} + \chi_{I4})\, \delta_-\epsilon_I \,,
\end{equation}
where $\chi_{II} =1$ and $\chi_{I\neq J}=0$ with $I,J$ labelling contour segments.
We can interpret the first term as capturing the effect of the difference field $\epsilon_{\dif(1,3)}(T)$ and the second term as capturing $\epsilon_{\dif(2,4)}(0)$, which appears in \eqref{eq:OTOCschw}.
This is illustrated in the figures above. 
The reason why the action is not zero can be found in the boundary conditions at the contour turning points, where $\delta\epsilon^\text{otoc}_I$ exhibits discontinuities.

We can derive an action, which describes the effective interaction between these modes along the contour (in particular on the segment $I=3$). 
Evaluating the quadratic action of Schwarzian fluctuations on $\delta\epsilon^\text{otoc}_I$ yields:
\begin{equation}
    iS_{\text{quad}}[\delta \epsilon^\text{otoc}] = 2iC\, \e^{-T} \, X^+ X^- \,.
\end{equation}
The coefficient of this `eikonal' action, $C\e^{-T}$, is ${\cal O}(1)$ at times of order the scrambling time. The path integral over $X^\pm$ must then be performed without resorting to the saddle point approximation.

Evaluating the bilocal operator \eqref{eq:Blinearized} on the configuration \eqref{eq:epsotoc} yields
\begin{equation}
    \begin{split}
    \left\langle{\cal B}^{\Delta_W}_{1,3}(T,T) \, {\cal B}^{\Delta_V}_{2,4}(0,0) \right\rangle_\beta
    = 1 - \frac{\Delta_V \Delta_W}{\sin^2 \varepsilon} \, \langle X^+ X^- \rangle_\beta 
    = 
    1-\frac{\Delta_V \Delta_W }{2iC\, \sin^2\varepsilon} \,\e^{T}\,.
    \end{split}
\end{equation}

Of course, there are other modes contributing to the OTO path integral. Our claim is merely that $\delta \epsilon_I^\text{otoc}$ gives the largest contribution for a large time separation $T$. Other `gravitational' (Schwarzian) contributions will be exponentially suppressed at times $t \sim t_\star \equiv \log N$.

\subsubsection{Gravity and strings}

In preparation for our discussion of scramblons for non-maximal chaos, we briefly review some aspects of the eikonal phase in holography and in string theory. We closely follow \cite{Shenker:2014cwa}.

Consider now a more generic case (allowing for higher dimensions) and assume that the theory in question has a semiclassical gravity dual. The thermofield double state corresponds to an eternal black hole geometry and the OTOC \eqref{eq:otocInOut} can be represented geometrically as the overlap of two states:
\begin{equation*}
\includegraphics[width=.5\textwidth]{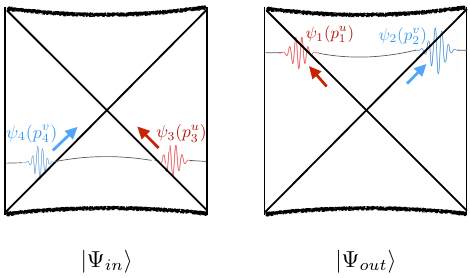}
\end{equation*}
The effect of each of the four operator insertions is approximated by a gravitational null shockwave \cite{Aichelburg,Dray:1984ha,Roberts:2014isa}. On the initial and final time slices we write the corresponding bulk field excitations as wave packets dressed to the boundary via bulk-to-boundary propagators:
\begin{equation}
    \begin{split}
        |\Psi_{\text{in}}\rangle
        &= {\color{RoyalBlue}W_L} {\color{Maroon}V_R} |\text{TFD}\rangle = \int  \d p_3^u \d p_4^v \; {\color{RoyalBlue}\psi_4(p_4^v)} {\color{Maroon} \psi_3(p_3^u)} \, |p_3^u,p_4^v \rangle_{\text{in}} \,,
        \\
        |\Psi_{\text{out}}\rangle
        &= {\color{Maroon}V_L} {\color{RoyalBlue}W_R} |\text{TFD}\rangle = \int  \d p_1^u \d p_2^v \; {\color{Maroon}\psi_1(p_1^u)} {\color{RoyalBlue} \psi_2(p_2^v)} \, |p_1^u,p_2^v \rangle_{\text{out}} \,.
    \end{split}
\end{equation}
In Einstein gravity, the wave packets are constructed from bulk-to-boundary propagators, for example:
\begin{equation}
    \psi_{1,V}(p_1^u) = \int \d v \; \e^{2i v p_1^u} \, G_{b\partial}(0,v;t_1) \,,\qquad G_{b\partial}(u,v;t) \equiv \langle \phi_V(u,v) V(t)^\dagger \rangle\,.
\end{equation}
If the time separation between boundary insertions is large, and hence the relative boost is large, we can approximate $p_1^u \approx p_3^u$ and $p_2^v \approx p_4^v$. In this approximation, assuming the 2-to-2 scattering process is elastic, we furthermore have
\begin{equation}
    |p_1^u,p_2^v \rangle_{\text{out}} \approx \e^{i\delta(s,b)} |p_1^u,p_2^v \rangle_{\text{in}} \,,\qquad s = 4 p_1^u p_2^v \,,\;\; b = |x_1-x_2|\,.
\end{equation}

The OTOC can now be computed as the overlap
\begin{equation}
\label{eq:otocgrav}
\text{OTOC} = \langle \Psi_{\text{out}} | \Psi_{\text{in}} \rangle 
= \int \d p^u \d p^v \; {\color{Maroon}\psi_1(p^u) \psi_3(p^u)} \, {\color{RoyalBlue} \psi_2(p^v) \psi_4(p^v) } \; \e^{i\delta(s,b)} \,.
\end{equation}
In Einstein gravity, the scattering process is dominated by graviton ladder diagrams. The {\it eikonal phase} $\delta(s,b)$ is proportional to the 2-to-2 scattering amplitude involving a single graviton exchange. It exponentiates in the high energy limit as shown above, and takes the schematic form
\begin{equation}
\label{eq:deltagrav}
 \delta_\text{grav}(s,b) \propto G_N \, s\, \e^{T-\mu b}\,,
\end{equation}
where $\mu$ is a parameter characterizing the black hole.

We can now use the effective description of scrambling to make an aspect of holography quite manifest. In fact, the gravitational expression \eqref{eq:otocgrav} is related to the boundary description derived from the Schwarzian action by a Fourier transform of wave packets with aligned momenta:
\begin{equation}
\begin{split}
\int \frac{\d p^u}{p^u} \;  {\color{Maroon}\psi_{1,V}(p^u) \psi_{3,V}(p^u)} \, \e^{-ip^u \, X^+} &\propto \left[ \frac{1}{1-\frac{i}{2\sin \varepsilon} \, X^+} \right]^{2\Delta_V}\,, \\
\int \frac{\d p^v}{p^v} \;  {\color{RoyalBlue}\psi_{2,W}(p^v) \psi_{4,W}(p^v)} \, \e^{-ip^v\, X^-} &\propto \left[ \frac{1}{1-\frac{i}{2\sin \varepsilon} \, X^-} \right]^{2\Delta_W}  \,.
\end{split}
\end{equation}
The expressions on the right are precisely the two-point functions of $V$ and $W$ in the presence of sources for the scramblon modes excited by the respective other operator. Indeed, these expressions coincide with \eqref{eq:Bschwarzian} when $f_I$ and $f_J$ are suitable non-linear transformations corresponding to the soft modes $\delta_\pm \epsilon_{I}$.

\subparagraph{Stringy effects:} Corrections to the eikonal phase due to the exchange of a string were studied in \cite{Shenker:2014cwa}. They found that the phase \eqref{eq:deltagrav} is replaced by an expression of the following form:
\begin{equation}
\label{eq:deltastringy}
 \delta_\text{stringy} \propto G_N \, \int \d^{\D-1}p\;\frac{\e^{ip\cdot x_{12}}}{p^2+\mu^2} \, \big(-i\alpha' \, s \, \e^T \big)^{J(p)-1}\,,  
\end{equation}
with 
\begin{equation}
    J(p) = 2 - \alpha' \, \frac{k^2+\mu^2}{2r_0^2} + \ldots\,.
\end{equation}
Here, $J(p)<2$ should be understood as the Regge trajectory of the string parameterized by transverse momentum. The momentum integral is dominated by the {\it graviton pole} $p^2 = -\mu^2$ when the impact parameter $b=|x_{12}|$ is large. In this case we have the simplification 
\begin{equation}
\delta_\text{stringy}(b \gg T) \approx \delta_\text{grav} \,.
\end{equation}
On the other hand, for small impact parameter, stringy effects are important and the integral is no longer dominated by the gravitational contribution. Instead, the saddle point involving $J(p)$ dominates and yields
\begin{equation}
\delta_\text{stringy}(b \ll T) \approx G_N \big( -i\alpha' \, s \, \e^{T-\mu b} \big)^{1- \alpha' \mu^2/2r_0^2} \times \ldots
\end{equation}

This result was obtained by working to first order in $\alpha'$. More generally, at finite string length, further modifications might be expected in the bulk computation of the OTOC, \eqref{eq:otocgrav}: first, the wave functions (bulk-boundary propagators) $\psi_i$ may be modified from their conformally invariant `gravitational' form. Second, it might be possible that there is an amplitude for {\it multi-string exchanges} \cite{Choi:2023mab}; here, we do not mean the exponentiation of single-string exchanges (which the eikonal phase already captures), but rather the exchange of multiple strings from a single vertex. These effects have not been studied in the bulk theory, but we will see that they appear to play a role indeed in certain boundary models that capture finite coupling effects beyond the Schwarzian approximation.

\subsubsection{Seeing strings from chaos}

The effective description of scramblons living and interacting on the OTO contour applies qualitatively almost unchanged in models with non-maximal chaos. We will review below that finite coupling effects in a spatially extended chaotic system can lead to an OTOC that strikingly resembles and generalizes the string theory prediction discussed above. 

We will still discuss the example of the SYK model. However, we shall now be interested in the operator spectrum beyond the Schwarzian mode. At finite $\beta J$, the Schwarzian action is no longer parametrically separated, and other modes will contribute to four-point functions. 

In order to also discuss spatial dependence, consider a chain of SYK models with nearest-neighbor interactions along the chain. The disorder averaged model turns out to be described by a chain of coupled Liouville-like theories \cite{Choi:2020tdj}:\footnote{The parameter $C \equiv \frac{N}{q^2}$ is large, i.e., $1 \ll q^2 \ll N$. It arises from the microscopic description where each site has $N$ Majorana fermions, coupled through Gau\ss ian random variables with $q$ indices.}
\begin{equation}
\label{eq:Schain}
\begin{split}
    iS=\frac{C}{4} \sum_{x=0}^{M-1}&  \int_0^{4T+2\pi} \d s_1 \d s_2\, \frac{\d t_I}{\d s_1} \frac{\d t_J}{\d s_2} \\&\times \left[  \frac{1}{4} \partial_{t_I} g_{{\scriptscriptstyle IJ},x} \,\partial_{t_J} g_{{\scriptscriptstyle IJ},x} -\mathcal{J}_0^2 \,\e^{g_{IJ,x}} -\mathcal{J}_1^2 \,\e^{\frac{1}{2}(g_{IJ,x}+g_{IJ,x+1})} \right]\,,
\end{split}
\end{equation}
where $g_{{\scriptscriptstyle IJ},x}(t_I,t_J)$ is bilocal field with a discrete spatial label $x \in \{0, \ldots, M-1\}$ indicating a distance along the chain. The coupling ${\cal J}_0^2$ measures an on-site interaction, while ${\cal J}_1^2$ parametrizes the interactions between neighboring sites. Two important parameters of the model are $0\leq v <1$ and $1<h(p)\leq 2$, defined by 
\begin{equation}
 {\cal J}^2 = \frac{v^2}{4 \cos^2 \left( \frac{\pi v}{2} \right)} \,, \qquad 
  \frac{h(h-1)}{2} = 1 + \frac{\gamma}{2} \, [\cos(p) - 1] \,,
\end{equation}
where $p$ is the spatial momentum along the chain and $\gamma \equiv {\cal J}_1^2/({\cal J}_0^2+{\cal J}_1^2)$. The Lyapunov exponent will turn out to be $\kappa(p) = v(h(p)-1)$ with $\kappa(0) = v$.

The thermal saddle point (for $\beta = 2\pi$) is translation invariant and takes the form
\begin{equation}
\label{eq:saddleLargeq}
\e^{g_{IJ,p}(t_I,t_J)} = \frac{\cos^2 \left( \frac{\pi v}{2} \right)}{\cos^2\left(\frac{v}{2}\left(\pi-i(t_I-t_J)\right)\right)}\,.
\end{equation}

Note that in this model there is no obvious $\text{SL}(2,\mathbb{R})$ symmetry that governs the dynamics. 
Nevertheless, studying the quadratic action of fluctuations around the above saddle point reveals the existence of zero modes $\delta_\pm g_{{\scriptscriptstyle IJ},p}$ with exponential time dependence. As before, there is a retarded and an advanced mode, which are excited by the past and future operator insertions, respectively. The OTOC is computed by the following combination:
\begin{equation}
\delta g_{{\scriptscriptstyle IJ},p}^\text{otoc}(t_I,t_J)=X^+(p)\,\delta_+ g_{{\scriptscriptstyle IJ},p}(t_I,t_J)+X^-(p)\,\delta_- g_{{\scriptscriptstyle IJ},p}(t_I,t_J)\,,
\end{equation}
where
\begin{equation}
\label{eq:deltagCHAIN}
\delta_\pm g_{{\scriptscriptstyle IJ},p}(t_I,t_J)=A_{\pm,{\scriptscriptstyle IJ}}(p)\left[\frac{\e^{\pm v ((1\mp 1)T-t_I-t_J)/2}}{ \cos\left({\frac{v}{2}}\left(\pi-i(t_I-t_J)\right)\right)}\right]^{h(p)-1}  \,, 
\end{equation}
where $A_{\pm, IJ}(p)=A_{\pm, JI}(p)$  are given by 
\begin{equation}
\begin{split}
\\&A_{+,IJ}(p) =\begin{sqcases}\e^{-\frac{i\pi }{2} \kappa(p)} &\quad(I,J)=(2,1),(3,1)\,,
\\
\e^{-\frac{3i\pi}{2} \kappa(p)} &\quad(I,J)=(4,2),(5,2),(4,3),(5,3)\,,
\\0 &\quad 	\text{otherwise}\,,
\end{sqcases}
\\& A_{-,IJ}(p)=\begin{sqcases}-\e^{\frac{3i\pi}{2}\kappa(p) } &\quad(I,J)=(3,1),(4,1),(3,2),(4,2)\,,
\\
-\e^{\frac{5i\pi}{2}\kappa(p)}&\quad(I,J)=(5,3),(5,4)\,,
\\0 &\quad \text{otherwise}	\,.
\end{sqcases}
\end{split}
\end{equation}
These normalization factors are `locally' arbitrary on each contour segment, but must be chosen in such a way along the contour turning points that $ \delta_\pm g_{{\scriptscriptstyle IJ},\,p}(s_1,s_2)$ satisfy the KMS conditions \begin{equation}
g_{{\scriptscriptstyle 5J},\,p}(4T+2\pi,s_2) = g_{{\scriptscriptstyle 1J},\,p}(0,s_2) \,,\qquad g_{{\scriptscriptstyle I5},\,p}(s_1,4T+2\pi) = g_{{\scriptscriptstyle I1},\,p}(s_1,0)
\,.
\end{equation} 

As in the Schwarzian theory, one finds an effective action for the soft modes $X^\pm(p)$:\footnote{For completeness \cite{Choi:2023mab}: $K(x) =  \frac{1}{2\sqrt{\pi}}\,\cos \left( \pi x \right) \Gamma \left( x+1 \right) \Gamma \left(\frac{1}{2} -x\right)$.}
\begin{equation} \label{EikonalChain}
\begin{split}
iS_{\text{quad}}[\delta g_{{\scriptscriptstyle IJ},p}^\text{otoc}] &= -iC \int_{-\pi}^\pi \frac{\d p}{2\pi} \; K\!\left( \frac{\kappa}{v}\right) \;{\color{RoyalBlue}\cos\left( \frac{\pi \kappa}{2} \right) \,\e^{\frac{i \pi}{2}(\kappa-1)-\kappa T} }\, X^+(p)X^-(-p)\,.
\end{split}
\end{equation}
We have highlighted the most important factors, which determine the properties of the propagator $\langle X^+(p) X^-(-p)\rangle_\beta$.

The above formulas are sufficient to compute the OTOC to order $1/C$, which is given by
\begin{equation}
\label{eq:stringyInt}
\begin{split}
& \int \frac{\d p}{2\pi} \, \e^{ipx} \, \big( \Delta_V\,\delta_+ g_{42,p}(0,0) \big)\,\big( \Delta_W\,\delta_- g_{31,-p}(T,T) \big)\,\left\langle X^+(p) X^-(-p) \right\rangle \\
 &\qquad = -\frac{\Delta_V \Delta_W}{C}\int \frac{\d p}{2\pi} \, \frac{{\color{RoyalBlue}\e^{ipx + \kappa\left(-\frac{i\pi}{2}+T\right)}}}{K\big(\frac{\kappa}{v}\big) \,{\color{RoyalBlue} \cos \big( \frac{\pi \kappa}{2} \big)}} \left[ \frac{1}{\cos \big( \frac{\pi v}{2} \big) }\right]^{{\frac{2\kappa}{v}}}\,.
\end{split}
\end{equation}
We refer to \cite{Choi:2023mab} for details, but only repeat the following points, which are completely analogous to the mechanism that applied for the stringy eikonal scattering phase \eqref{eq:deltastringy}:
\begin{itemize}[label=$\diamond$]
\item For $\frac{x}{T}$ is sufficiently small the integral is dominated by a saddle point that extremizes the exponential. The Lyapunov exponent in this case turns out to be $v$ and the OTOC decays exponentially in all spatial directions. 
\item On the other hand, if $\frac{x}{T}$ is large, then the integral is dominated by the pole at $\kappa = 1$; this pole can be interpreted as due to a ``stress tensor'' exchange. The resulting Lyapunov exponent is maximal.
\end{itemize}
In conclusion, the theory of scramblons in the large $q$ SYK chain has a close structural similarity with eikonal scattering in perturbative string theory. In this sense, we see the emergence of a stringy bulk description from the effective theory of chaos. We also point out that there appear to be two competing effects leading to a sub-maximal chaos exponent: on the one hand, the genuinely stringy effects  that control the momentum-dependent integral \eqref{eq:stringyInt}, on the other hand the presence of the $v$ parameter, which ``stretches" the thermal circle (e.g., in \eqref{eq:saddleLargeq}). It would be interesting to understand how general this distinction is and if the latter effect has an interpretation in terms of the bulk S-matrix.

\section*{Appendix: Schwinger--Keldysh supersymmetry}\label{sec:sksusy}

 It turns out to be rather useful to formulate Schwinger--Keldysh unitarity as a symmetry statement. Clearly, the doubled theory has a topological flavor to it, where the difference operators are sensitive only to `global' properties. More precisely, we can introduce a {\it Schwinger--Keldysh BRST cohomology} where the difference operators are BRST-exact under CPT-conjugate charges ${\cal Q}_{SK}$ and $\overline{\cal Q}_{SK}$. For every (say, bosonic) operator ${\cal O}$ we introduce a quadruplet consisting of the bosonic operators ${\cal O}_\av$ and ${\cal O}_\dif$, as well as a ghost ${\cal O}_\G$ and an anti-ghost ${\cal O}_\Gb$. We collect these into a superfield
 \begin{equation}
     \mathring{{\cal O}} = {\cal O}_\av + \theta \, {\cal O}_\Gb + \bar\theta \, {\cal O}_\G + \bar\theta\theta \, {\cal O}_\dif \,,
 \end{equation}
 such that translation invariance in the Grassmann-odd directions implements the SK BRST cohomology. For example, ${\cal O}_\dif = \partial_\theta \partial_{\bar\theta} \mathring{{\cal O}}$ is BRST-exact. This is as desired, but it is a little too naive: the topological nature of ${\cal O}_\dif$ should manifest itself if the operator is future-most within a correlation function. To implement this, we need to fix the behavior of the initial and final states with respect to the BRST charges. One finds that the following rules lead to a consistent picture:
 \begin{itemize}
 \item[$(i)$] The {\it final state} (future turning point) is super-translation invariant, i.e., it is annihilated by both BRST charges. This is motivated by the fact that the future turning point in the SK path integral projects onto a maximally entangled state (without any ghosts). 
 \item[$(ii)$] The {\it initial state}, being an arbitrary density matrix, generically carries some ghost zero modes. These can, without loss of generality, be parameterized as a background superfield 
 \begin{equation}
  \mathring{{\cal O}}_0 \equiv 1 + \theta_0 \, \bar{g}_0 + \bar{\theta}_0 \, g_0 + \bar\theta_0\theta_0 \, d_0 \,,
 \end{equation}
 which accompanies the initial density matrix in any correlation function, i.e., we effectively replace $\rho_0 \rightarrow \mathring{\rho}_0 \equiv \mathring{{\cal O}}_0\, \rho_0$. \end{itemize}
These conditions can be conveniently formulated in the quadrupled Hilbert space with ghost components: $\mathring{{\cal H}}={\cal H}_1 \otimes {\cal H}_2 \otimes {\cal H}_G \otimes {\cal H}_{\bar G}$. All operators (including the density matrix) get uplifted to quadruplets in this Hilbert space, and the future turning point is associated with a final state $|\mathring{f}\rangle \in \mathring{{\cal H}}$. The conditions above then amount to demanding:
\begin{equation}
\begin{split}
  (i)& \qquad {\cal Q}_{SK} |\mathring{f}\rangle = \overline{{\cal Q}}_{SK} |\mathring{f}\rangle = 0 \,,\\
  (ii)& \qquad {\cal Q}_{SK} |\mathring{\rho}_0\rangle = g_0  |\mathring{\rho}_0\rangle\,,\quad
   \overline{{\cal Q}}_{SK} |\mathring{\rho}_0\rangle = \bar{g}_0  |\mathring{\rho}_0\rangle \,.
\end{split}
\end{equation}
Showing consistency of this ansatz amounts to showing that every newly introduced correlator can be consistently fixed in terms of the standard real-time correlators present in the original SK path integral. This will ensure that there is no net new information and the largest time equation remains upheld. The newly introduced correlators include: $(i)$ those involving any of the SK ghosts, and $(ii)$ those involving the background field $\mathring{{\cal O}}_0$. It was shown in \cite{Haehl:2016pec} that super-translation invariance fixes these consistently will all constraints.

\newpage
\bibliographystyle{jhep}
\bibliography{arxiv.bbl}

\end{document}